%% file: main.tex
\documentclass[ nofootinbib, aps, prb, 10pt]{revtex4-2}
\usepackage{mystyle}

\begin{document}

\subfile{sections/A_title_abstract_luca.tex}

\subfile{sections/B_intro_luca.tex}

\subfile{sections/C_gauge_luca.tex}

\subfile{sections/D_TN_luca.tex}

\subfile{sections/E_ansatz_luca.tex}

\subfile{sections/F_doublesym_luca.tex}


\subfile{sections/GA_generalisation_luca.tex}

\subfile{sections/H_conclusion_luca.tex}

\input{sections/I_aknowledgments_luca.tex}
\bibliography{lgt_peps,bib}

\end{document}

%% file: sections/A_title_abstract_luca.tex
\title{A tensor network formulation of Lattice Gauge Theories based only on symmetric tensors}
\author{Manu Canals}
\email{manu.canals@icfo.eu} 
\affiliation{Arnold Sommerfeld Center for Theoretical Physics, Center for NanoScience,
and Munich Center for Quantum Science and Technology,
Ludwig-Maximilians-Universität München, 80333 Munich, Germany}
\author{Natalia Chepiga}
\email{N.Chepiga@tudelft.nl}
\affiliation{Kavli Institute of Nanoscience, Delft University of Technology, Lorentzweg 1, 2628 CJ Delft, The Netherlands}
\author{Luca Tagliacozzo}
\email{luca.tagliacozzo@iff.csic.es} 
\affiliation{Institute of Fundamental Physics IFF-CSIC, Calle Serrano 113b, Madrid 28006, Spain}

\begin{abstract}
{\bf Abstract:} The Lattice Gauge Theory  Hilbert space is divided into gauge-invariant sectors selected by the background charges.  Such a projector can be directly embedded in a tensor network ansatz for gauge-invariant states as originally discussed in [Phys. Rev. B 83, 115127 (2011)] and in [Phys. Rev. X 4, 041024 (2014)] in the context of Projected Entangled Pair States (PEPS). The original ansatz is based on sparse tensors, though parts of them are not explicitly symmetric, and thus their actual implementation in numerical simulations has been hindered by the complexity of developing ad hoc libraries.
Here we provide a new PEPS tensor network formulation of gauge-invariant theories purely based  on symmetric tensors. The new formulation can be implemented in numerical simulation using available state-of-the-art tensor network libraries but also holds interest from a purely theoretical perspective since it requires embedding the original gauge theory with gauge symmetry $G$ into an enlarged globally symmetric theory with symmetry $G{\times}G$. We also  revisit the original TN ansatz in the modern landscape of i) duality transformations between gauge and spin systems, ii) finite depth quantum circuits followed by measurements that allow generating topologically ordered states, and iii) Clifford enhanced tensor networks. Our new formulation should  provide new insights in those landscape. 

\end{abstract}

\maketitle

%% file: sections/B_intro_luca.tex
\section{Introduction}
\label{sec:intro}

Many-body quantum systems still defeat our understanding in situations where their constituents interact strongly. As a result, often we do not understand their emerging properties from first principles. For example, we are struggling to describe the phase diagram of nuclear matter at finite quark density, relevant for understanding the physics of quark gluon plasmas present in the interior of stars and reproduced experimentally in heavy ion collisions.

The standard lore is that our inability arises from the exponential increase in the complexity of describing many-body quantum states as the number of their constituents increases. However, recent developments have shown that most of the relevant quantum states can be expressed using Tensor Networks (TN) \cite{poulin2011}, which require only a polynomial number of parameters in relation to the number of constituents. Unfortunately, not all of these Tensor Networks are exactly contractible. Nevertheless, it is believed that in many scenarios, approximate contraction schemes yield good results, providing at least a qualitative understanding and, in most cases, a good quantitative picture of the many-body phase diagram \cite{orus2014,bridgeman2017,ran2020,banuls2023}.

Gauge theories and their lattice formulation constitute a specific class of many-body quantum systems \cite{kogut1975,creutz1977,kogut1979a}. In these systems, the "Gauss law" divides the Hilbert space into different sectors that are not mixed by the dynamics, a property that can have profound consequences for the system's ergodicity as shown for examples in specific quench protocols \cite{chanda2020}.

Since TN mostly provide approximate descriptions of quantum states, it is desirable to separate the effects of the approximations from the existence of distinct gauge sectors. Furthermore, it is important to have access to the physics within a specific gauge sector, even if that sector does not have the lowest variational energy.

As a result, it is desirable to encode the gauge-invariant sector directly into the Tensor Network ansatz. Such a program was initiated in \cite{tagliacozzo2011}, unveiling a deep connection between the use of gauge-invariant (and, more generally, symmetric) Tensor Networks and duality transformations. In particular, it was shown that a unitary map could be constructed by concatenating elementary tensors from the Clifford group, which locally implement a duality transformation between the gauge-invariant theory and a globally symmetric theory. This approach allows us to interpret the gauge-invariant Tensor Network as a local version of the well-known duality between spin and gauge systems \cite{savit1980}.

In modern terms, such results could also be interpreted as a Clifford enhanced tensor network, since the Clifford part of the network addresses part of the entanglement in the gauge-invariant wavefunction, while the rest should address the "magic" part of the wavefunction's entanglement \cite{mello2024,fux2024}.

Originally, gauge-invariant tensor network were formulated for discrete groups within a renormalization group framework (leading to a MERA network \cite{vidal2007}), but subsequent work \cite{tagliacozzo2014} generalized these results to arbitrary groups. This new version can be interpreted, following the philosophy of \cite{tagliacozzo2011}, as an alternative duality transformation that, rather than focusing on mapping local operators to local operators, focuses on having the most shallow form of the unitary transformation implementing the duality. As we will review, this indeed is implemented by finite-depth unitary circuit followed by measurements, something that consequently maps local operators to Tensor Product Operators (TPOs), the generalization of matrix product operators to arbitrary spatial dimensions.  Alternatively one could say that in such a formulation, local operators should be dressed by global measurements, and thus they don't exist in a first instance. Regardless of the perspective,  the projector onto the gauge-invariant subspace is encoded by a TPO, a finite-depth unitary circuit followed by measurements. As a result, these ansatz provide  early examples of what has been  later understood as the possibility to generate topological order from a finite-depth circuit enhanced with measurements, as discussed in \cite{sahay2024}. The resulting Tensor Network encodes gauge-invariant states as a Projected Entangled Pair State (PEPS) or a Tensor Product State (TPS) \cite{nishio2004,verstraete2004a}.

In parallel, numerous numerical results have been obtained in one dimension \cite{banuls2013,banuls2015,buyens2014,kuhn2015,buyens2016,pichler2016}. Similar Tensor Network approaches have also been proposed for two and higher dimensions \cite{rico2014,silvi2014,haegeman2015,zohar2016,zohar2019,robaina2021}, as reviewed in several sources \cite{banuls2020,banuls2023}.

Despite the theoretical relevance of the original formulations, actual numerical simulations have been hindered by the complexity of such ansatz, since they include different types of elementary tensors, which are sparse but not necessarily symmetric and hence are not supported by standard tensor network libraries but rather require ad-hoc complex coding in order to exploit their sparseness.

Here, we present a novel formulation of gauge-invariant Tensor Networks that enables the description of Lattice Gauge Theories using only explicitly symmetric elementary tensors \cite{singh2010,singh2011}. These tensors enhance the performance of numerical algorithms and can be simulated using standard Tensor Network libraries \cite{hauschild2018,hauschild2024,fishman2022,weichselbaum2024,rams2024}.




 Our strategy is to double the symmetry of individual constituents, and embed the original Hilbert space in a subsector of a larger space, mapping \textit{one} local symmetry to \textit{two} global symmetries, as discussed in Sec.~\ref{sec:doublesym}. This procedure is completely general, and provides a new relation between lattice gauge theory and globally invariant systems.
 
 The paper is organized as follows:  in Sec.~\ref{sec:gauge} we review the lattice formulation of gauge theories, in Sec.~\ref{sec:TN} we review the basic notion of TN and symmetric TN, in Sec.~\ref{sec:ansatz} we review the construction of gauge-invariant Tensor Networks and their connections to i) the preparation of topological state using finite-depth circuits plus measurements, and ii) the Clifford enhanced TN networks. In Sec.~\ref{sec:doublesym} we provide a warm-up example on how to use a TN ansatz for gauge-invariant states, built from elementary tensors which are symmetric. In Sec.~\ref{sec:general} we present the generalization to arbitrary \zn groups. We finish with conclusions and outlooks.

%% file: sections/C_gauge_luca.tex
\section{Gauge Theory}
\label{sec:gauge}
In this section, we focus on an oriented \rrr{$D-$dimensional square} Lattice Gauge Theories in the Hamiltonian formulation \cite{kogut_hamiltonian_1975,creutz_gauge_1977}. For simplicity, we will focus on Abelian gauge theories with gauge group \zn and on $D=2$. This group is the group of addition modulo $N$, or equivalently of the multiplication of the $N_{\textrm{th}}$ roots of the identity, $h_n = \exp(-i \frac{2 \pi n}{N})$ with $n=0,\hdots, N-1$. Lattice gauge theories are traditionally defined in the Hilbert space of the group algebra, where to every element of the group one associates an \rrr{orthonormal} vector, i.e. \rrr{$g\mapsto\ket{g}$ where $\braket{g|f}=\delta_{g,f},\forall g,f\in\zn$}, though different choices are possible \cite{tagliacozzo2014}. In such a Hilbert space, \rrr{$\mathcal{S}$}, one can define the rotation of a state by an arbitrary element of the group by using the corresponding regular matrix representation $R_\text{\rrr{$\mathcal{S}$}}(g)$.

In this scenario, the matter field \rrr{on an oriented $D$-dimensional square lattice $\Lambda$} is described by constituents located at the vertex sites \rrr{$i$}, whose set is denoted by $\mathcal{V}.$ \rrr{On the other hand}, gauge bosons are described by constituents defined on link sites $l$ of $\Lambda$, whose set is $\mathcal{L}$. Elementary closed loops on $\Lambda$ are plaquettes, and  their set is $\mathcal{P}$. 
It is also useful to introduce the concept of a \textit{star} \rrr{associated with each vertex $\rrr{i}\in\mathcal{V}$}. A star is defined as the set \rrr{of constitutents} on the links starting from or ending at vertex $\rrr{i}$, including the matter constituent at that vertex. \rrr{In the global Hilbert space $\bigotimes_{s\in \Lambda}\mathcal{S}^{[s]}$}, we can thus define a local operator $A_\rrr{i}(g)$ at each vertex $\rrr{i}\in\mathcal{V}$, \rrr{acting on the associated star as}
\begin{equation}
A_{\rrr{\rrr{i}}}(\rrr{g}) =
R_\rrr{i}(\rrr{g})
\prod_{l^+\in \rrr{\rrr{i}}}
R_{l^+}(\rrr{g})
\prod_{l^-\in \rrr{\rrr{i}}}
R_{l^-}^{\dagger}(\rrr{g}),
\label{eq:A_s}
\end{equation}
where $l^+$ denotes the links entering the vertex $\rrr{i}$, and $l^-$ those exiting it.
Given such a set of local operators, a gauge theory is defined by any choice of a Hamiltonian $\ham$ that commutes with $A_{\rrr{\rrr{i}}}(\rrr{g})$ for all \rrr{possible choices of} $\rrr{i}\in\mathcal{V}$ and $\rrr{g}\in\zn$. \rrr{Explicitly,}
\begin{equation}
 [\ham,A_{\rrr{\rrr{i}}}(\rrr{g})] =0,\,\quad\forall\rrr{\rrr{i}}\in\rrr{\mathcal{V}},\, \quad \forall \rrr{g} \in \rrr{\zn}.  
\end{equation}
Such Hamiltonians can vary widely, and a specific choice is given by the Kogut-Susskind Hamiltonian \cite{kogut_introduction_1979} defined as 
\begin{equation}
 \ham = -{1}/{\rrr{g^2_{KS}}}\left( \sum_{p\in\mathcal{P}}B_p+h.c\right) - \left(\sum_{l\in\mathcal{L}}R_l(h_1)+h.c.\right)   \label{eq:KSham}
 \end{equation}
where 
\begin{equation}
B_p = \prod_{l^+\in p} \Omega_{l^+} \prod_{l^-\in p} \Omega^{\dagger}_{l^-}.   \label{eq:Bp}
\end{equation}
Here, the $l^{+}$ links are those that, following a counter-clock wise orientation of the plaquette, are visited along the same orientation than $\Lambda$'s, while $l^-$ are those oriented opposedly. $\Omega$ is the $N\times N$ diagonal matrix containing the phases $h_n$ along the diagonal,
\begin{equation}
\Omega =\sum_n h_n \ket{n}\bra{n}\ .
\end{equation}

Given that, using the Abelian nature of the group $A_{\rrr{\rrr{i}}}(h_n) = A_\rrr{i}(h_1)^n$, and that each of these generators commute with the Hamiltonian, we can divide the Hilbert space in different gauge sectors. These sectors are characterized by the choice of the eigenvalue of $A_{\rrr{\rrr{i}}}(h_1)$ \rrr{at each vertex $\rrr{i}\in\mathcal{V}$}. \rrr{The set of possible} eigenvalues are just the $h_n$ phases. Once we start from a state in a given sector, given that all $A_{\rrr{\rrr{i}}}(\rrr{h_1})$ commute with the Hamiltonian, we will never leave such sector if the dynamics are generated by $\ham$. 
\rrr{We can thus specify a given sector $K$ of the gauge theory by choosing one phase $\exp\left(-i\frac{2\pi}{N}g_i\right)$ for each vertex $i$, where $g_i\in\{0,\dots,N-1\}$. }States in a given  sector fulfill
\begin{equation}
    K(\{\rrr{g_i}\})
    =
    \bigl\{
    \ket{\Psi}\ 
    \text{such that}
    \ A_{\rrr{\rrr{i}}}(h_1)\ket{\Psi}
    {=}
    \rrr{\exp\left(-i\frac{2\pi}{N}g_i\right)}\ket\Psi,
    \ \forall
    \rrr{\rrr{i}}{\in}\rrr{\mathcal{V}}
    \bigr\}.
    \label{eq:gauge:K}
\end{equation}
The choice of \rrr{$g_i\in\{0,\dots,N-1\}$} is attributed to the presence of a background charge \rrr{$g_i$} at site $\rrr{i}\in \mathcal{V}$, since the above local constraint described by each $A_\rrr{\rrr{i}}(h_1)$ can be interpreted as a lattice $\mathbb{Z}_N$ version of the Gauss law \cite{kogut_hamiltonian_1975,tagliacozzo_entanglement_2011}. The sector of the gauge theory associated with the absence of background charges is the sector where all $\rrr{g_i=0}$. 

We now provide an explicit example for of the simplest $\mathbb{Z}_2$ gauge theory, where \rrr{each constituent} Hilbert space $\rrr{\mathcal{S}}$ is described by spin $1/2$ particles on which we can define the standard Pauli matrices $\sigma_\rrr{x}, \sigma_\rrr{y}$ and $\sigma_\rrr{z}$ obeying the usual rules $\sigma_i\sigma_j=i\epsilon_{ijk}\sigma_k$, with $\epsilon$ the fully antisymmetric tensor with element $\epsilon_{123} =1$. In this case, $R_\rrr{\mathcal{S}}(h_1)=\sigma_x$ and  $\Omega_\rrr{\mathcal{S}} = \sigma_z$. We will also focus on the pure gauge theory where we have no matter degrees of freedom, and thus the $A_\rrr{\rrr{i}}(\rrr{g})$ is restricted to act on the links \rrr{of the associated} star.

In $D=1$, such theory is trivial if defined on an open chain, and just describes a single spin degree of freedom if defined on a ring encoding the $\mathbb{Z}_2$ flux piercing the ring. In $D=2$, the theory possesses two phases: a de-confined phase at weak coupling $\rrr{g_{KS}^2} \ll 1$ and a confined strong coupling phase for $\rrr{g_{KS}^2} \gg 1$\rrr{. T}he two phases are separated by a second order phase transition in the universality class of the 3D Ising model \cite{wegner_duality_1971}.

It is important to notice, that the constraints that fix a gauge sector $K(\{\rrr{g_i}\})$ spoil the tensor product structure of the Hilbert space. In the next section, we review how we can nevertheless use tensor networks to describe such states.

%% file: sections/D_TN_luca.tex
\section{Tensor Networks for specific gauge sectors}
\label{sec:TN}
Tensor networks describe the states of many-body quantum systems as the contraction of elementary tensors. It is customary to avoid explicitly writing tensor contractions when there are many tensors and instead use a diagrammatic notation. In this notation, tensors are represented by geometric shapes, with each open leg corresponding to one of their indices. Given that indices referring to bra and ket states rotate differently under a change of basis, we use arrows to distinguish them. For example, the elements of a rank-three tensor $T$ in a given basis are:
\begin{equation}
    T=\sum_{a,b,c}T_c^{ab}\ket{c}\bra{a}\bra{b}
    ,\quad
    T^{ab}_c \longleftrightarrow\  
    \input{editions/gen_compo_eg} .
\end{equation}

Tensors can also be invariant under specific symmetry actions.  Here we focus on a \zn symmetry. \rrr{\zn-}invariant tensors with total charge $a_\mathrm{tot}\in\{0,\dots,N-1\}$ fulfill 
the following restriction, given diagrammatically and analytically:
\begin{align}
    \input{editions/rotatedten_zn}
    \hspace{-4pt}
    =&
    \exp\left(-\frac{i a_{tot} 2\pi}{N}\right)\times
    \hspace{-12pt}
    \input{editions/ten_zn}\ ,\nonumber\\
    R_\rrr{1}(g)^{i'_1}_{i_1}\hdots
    R_\rrr{j}(g)^{i'_j}_{i_j}
    {R^{\dagger}_\rrr{j+1}(g)}^{o_\rrr{j+1}}_{o'_\rrr{j+1}}\hdots {R^{\dagger}_\rrr{J}(g)}^{o_\rrr{J}}_{o'_\rrr{J}}
    C ^{i_1\dots i_{\rrr{j}}}_{o_\rrr{j+1}\dots o_\rrr{J}} 
    =&
    \exp\left(-\frac{i a_{tot} 2\pi}{N}\right)
    C ^{i'_1\dots i'_\rrr{j}}_{o'_\rrr{j+1}\dots o'_\rrr{J}}\ ,
    \label{eq:inv_tensor}
\end{align}
where $R\rrr{_j(g)}$ are the unitaries representing the rotation under a given element \rrr{$g\in\zn$} \rrr{of a given leg $j$}. Such rotations can be diagonalized, and given that the irreducible representations of Abelian groups are all one dimensional, each diagonal element identifies an irreducible representation of the group. The number of eigenvectors that rotate with the same phase identify  the multiplicity of the specific irreducible representation. The eigenvalues $h_n = \exp\left(-i \frac{2\pi a_n}{N}\right)$ label the charge $a_n$ of the irreducible representation.

As a result, the Hilbert space spanned by each tensor's leg, \rrr{$\mathcal{S}_j$}, can be decomposed as a sum of a given set of irreducible representation, 
\begin{equation}
    \rrr{\mathcal{S}_j} =\bigoplus_{i=1}^N( a_i \otimes \alpha_i),
\end{equation}
where $a_i$ represent the charges of the irreducible representation and $\alpha_i$ their arbitrary multiplicity.  
Equation  \eqref{eq:inv_tensor} thus implies that symmetric tensors are a generalization of block-diagonal matrices. The non-zero blocks of the tensors are selected by a charge conservation requirement, giving rise to  \emph{sum-rules}. Blocks that do not satisfy the sum rule must be zero, as enforced by the symmetry constraints, namely:
\begin{equation}
   C =  \begin{cases}
        C^{(a_1;\alpha_1)\dots (a_\rrr{j};\alpha_\rrr{j})}_{(o_{\rrr{j+1}};\beta_\rrr{j+1})\dots (o_{\rrr{J}};\beta_\rrr{J})}
        & 
        \parbox[t]{.40\textwidth}{\raggedright
        if $a_1{+}{\dots}{+}a_\rrr{j}{-}o_{\rrr{j+1}}{-}{\dots}{-}o_{\rrr{J}}{=}a_\mathrm{tot}$ $(\textrm{mod}\  N)$
        }
        \\
        0 & \text{otherwise}
    \end{cases}\ .
\end{equation}

The above equation summarizes the fact that the fusion of the irreducible representation of each of the constituent should be the selected irreducible representation that defines the total charge to the tensor. If the tensor is invariant under the application of the unitaries from $G$, we can only select irreducible representations that fuse to the trivial one, meaning that $a_{\textrm{tot}}=0
$.

In the specific case of a $\mathbb{Z}_2$ tensor, given that each element is the inverse of itself, one can ignore orientation and the restrictions on non-zero blocks simplifies to a \ztwo sum-rule \rrr{--- notice the variation on the component's notation ---:}
\begin{gather}
    \input{editions/Ztwo_ten}
    =\begin{cases}
        C^{a_1\dots a_R}_{\alpha_1\dots\alpha_R} & 
        \parbox[t]{.16\textwidth}{\raggedleft
        if $a_1{+}{\dots}{+}a_R{=}a_\mathrm{tot}$ $\modtwo$
        }
        \\
        0 & \text{otherwise}
    \end{cases}
    \ .
    \def \myshift {80pt}
    \def \mybracelen {70pt}
    \hspace{-\myshift}
    \raisebox{17pt}{$\overbrace{\hspace{\mybracelen}}^\text{\ztwo sum-rule}$}
    \hspace{7pt}
    \label{eq:TN:Ccompo}
\end{gather}

In the same way that contracting standard tensors—one per constituent—allows one to express families of many-body quantum states, contracting symmetric tensors enables the expression of symmetric states, which have specific covariance properties under the rotations of the group $R_\rrr{i}(g), \forall i{\in}\mathcal{V}, \forall g{\in} \zn$. In the next section, we will show how this construction can be generalized to describe states in the gauge-invariant sectors $K(\{\rrr{g_i}\})$.

%% file: editions/gen_compo_eg.tex
\ensuremath{
\begin{matrix}
    \includegraphics[scale=1.1]{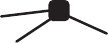}
\end{matrix}
\def \myx {67pt}
\def \myy {14pt}
\hspace{-\myx}
\raisebox{\myy}{
$\ket{a}$
}
\hspace{\myx}
\def \myx {80pt}
\def \myy {-17pt}
\hspace{-\myx}
\raisebox{\myy}{
$\ket{b}$
}
\hspace{\myx}
\def \myx {45pt}
\def \myy {-14pt}
\hspace{-\myx}
\raisebox{\myy}{
$\ket{c}$
}
\hspace{\myx}
\def \myx {90pt}
\def \myy {19pt}
\hspace{-\myx}
\raisebox{\myy}{
$T$
}
\hspace{\myx}
\def \myx {118pt}
\def \myy {5pt}
\def \myangle {8}
\hspace{-\myx}
\raisebox{\myy}{
\rotatebox[origin=c]{\myangle}{$
\begin{tikzpicture}[line width=1.5pt, color=gray!165]
    \draw [-{Classical TikZ Rightarrow[length=3.5pt]}](0,0) -- (0.05,0);
\end{tikzpicture}
$}}
\hspace{\myx}
\def \myx {123pt}
\def \myy {-4pt}
\def \myangle {35}
\hspace{-\myx}
\raisebox{\myy}{
\rotatebox[origin=c]{\myangle}{$
\begin{tikzpicture}[line width=1.5pt, color=gray!165]
    \draw [-{Classical TikZ Rightarrow[length=3.5pt]}](0,0) -- (0.05,0);
\end{tikzpicture}
$}}
\hspace{\myx}
\def \myx {104pt}
\def \myy {-1.2pt}
\def \myangle {-25}
\hspace{-\myx}
\raisebox{\myy}{
\rotatebox[origin=c]{\myangle}{$
\begin{tikzpicture}[line width=1.5pt, color=gray!165]
    \draw [-{Classical TikZ Rightarrow[length=3.5pt]}](0,0) -- (0.05,0);
\end{tikzpicture}
$}}
\hspace{\myx}
\hspace{-95pt}
}

%% file: editions/rotatedten_zn.tex
\ensuremath{
\begin{matrix}
    \includegraphics[scale=1.7]{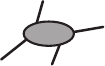}
\end{matrix}
\def \myx {100pt}
\def \myy {-3pt}
\hspace{-\myx}
\raisebox{\myy}{$
R_\rrr{1}\rrr{(g)}
$}
\hspace{\myx}
\hspace{-8.3pt}
\def \myx {93pt}
\def \myy {-11.5pt}
\hspace{-\myx}
\raisebox{\myy}{$
\begin{tikzpicture}[line width=1.5pt, color=gray!165]
    \filldraw (0,0) circle (1.5pt);
\end{tikzpicture}
$}
\hspace{\myx}
\def \myx {97pt}
\def \myy {-8pt}
\def \myangle {42}
\hspace{-\myx}
\raisebox{\myy}{
\rotatebox[origin=c]{\myangle}{$
\begin{tikzpicture}[line width=1.5pt, color=gray!165]
    \draw [-{Classical TikZ Rightarrow[length=3.5pt]}](0,0) -- (0.05,0);
\end{tikzpicture}
$}}
\hspace{\myx}
\def \myx {119pt}
\def \myy {-17pt}
\hspace{-\myx}
\raisebox{\myy}{
\rotatebox[origin=c]{\myangle}{$
\begin{tikzpicture}[line width=1.5pt, color=gray!165]
    \draw [-{Classical TikZ Rightarrow[length=3.5pt]}](0,0) -- (0.05,0);
\end{tikzpicture}
$}}
\hspace{\myx}
\def \myx {70pt}
\def \myy {-20pt}
\def \dx {-7pt}
\hspace{-\myx}
\hspace{\dx}
\raisebox{\myy}{$
R_\rrr{j}\rrr{(g)}
$}
\hspace{\myx}
\hspace{-7.7pt}
\hspace{-\dx}
\def \myx {102.5pt}
\def \myy {-16pt}
\hspace{-\myx}
\raisebox{\myy}{$
\begin{tikzpicture}[line width=1.5pt, color=gray!165]
    \filldraw (0,0) circle (1.5pt);
\end{tikzpicture}
$}
\hspace{\myx}
\def \myx {110pt}
\def \myy {-12.5pt}
\def \myangle {70}
\hspace{-\myx}
\raisebox{\myy}{
\rotatebox[origin=c]{\myangle}{$
\begin{tikzpicture}[line width=1.5pt, color=gray!165]
    \draw [-{Classical TikZ Rightarrow[length=3.5pt]}](0,0) -- (0.05,0);
\end{tikzpicture}
$}}
\hspace{\myx}
\def \myx {124.7pt}
\def \myy {-21.7pt}
\hspace{-\myx}
\raisebox{\myy}{
\rotatebox[origin=c]{\myangle}{$
\begin{tikzpicture}[line width=1.5pt, color=gray!165]
    \draw [-{Classical TikZ Rightarrow[length=3.5pt]}](0,0) -- (0.05,0);
\end{tikzpicture}
$}}
\hspace{\myx}
\def \myx {103pt}
\def \myy {-7pt}
\hspace{-\myx}
\raisebox{\myy}{$
R^\dagger_\rrr{j+1}\rrr{(g)}
$}
\hspace{4.4pt}
\hspace{\myx}
\def \myx {141pt}
\def \myy {4.7pt}
\hspace{-\myx}
\raisebox{\myy}{$
\begin{tikzpicture}[line width=1.5pt, color=gray!165]
    \filldraw (0,0) circle (1.5pt);
\end{tikzpicture}
$}
\hspace{\myx}
\def \myx {155.5pt}
\def \myy {2.4pt}
\def \myangle {11}
\hspace{-\myx}
\raisebox{\myy}{
\rotatebox[origin=c]{\myangle}{$
\begin{tikzpicture}[line width=1.5pt, color=gray!165]
    \draw [-{Classical TikZ Rightarrow[length=3.5pt]}](0,0) -- (0.05,0);
\end{tikzpicture}
$}}
\hspace{\myx}
\def \myx {151pt}
\def \myy {4.8pt}
\hspace{-\myx}
\raisebox{\myy}{
\rotatebox[origin=c]{\myangle}{$
\begin{tikzpicture}[line width=1.5pt, color=gray!165]
    \draw [-{Classical TikZ Rightarrow[length=3.5pt]}](0,0) -- (0.05,0);
\end{tikzpicture}
$}}
\hspace{\myx}
\def \myx {240pt}
\def \myy {18pt}
\hspace{-\myx}
\raisebox{\myy}{$
R_\rrr{J}^\dagger\rrr{(g)}
$}
\hspace{3.6pt}
\hspace{\myx}
\def \myx {233pt}
\def \myy {19.5pt}
\hspace{-\myx}
\raisebox{\myy}{$
\begin{tikzpicture}[line width=1.5pt, color=gray!165]
    \filldraw (0,0) circle (1.5pt);
\end{tikzpicture}
$}
\hspace{\myx}
\def \myx {241pt}
\def \myy {23pt}
\def \myangle {76}
\hspace{-\myx}
\raisebox{\myy}{
\rotatebox[origin=c]{\myangle}{$
\begin{tikzpicture}[line width=1.5pt, color=gray!165]
    \draw [-{Classical TikZ Rightarrow[length=3.5pt]}](0,0) -- (0.05,0);
\end{tikzpicture}
$}}
\hspace{\myx}
\def \myx {254.7pt}
\def \myy {13.5pt}
\hspace{-\myx}
\raisebox{\myy}{
\rotatebox[origin=c]{\myangle}{$
\begin{tikzpicture}[line width=1.5pt, color=gray!165]
    \draw [-{Classical TikZ Rightarrow[length=3.5pt]}](0,0) -- (0.05,0);
\end{tikzpicture}
$}}
\hspace{\myx}
\def \myx {245pt}
\def \myy {18pt}
\hspace{-\myx}
\raisebox{\myy}{$\dots$}
\hspace{\myx}
\def \myx {280pt}
\def \myy {-15pt}
\hspace{-\myx}
\raisebox{\myy}{$\dots$}
\hspace{\myx}
\def \myx {280pt}
\def \myy {0pt}
\hspace{-\myx}
\raisebox{\myy}{$\rrr{C}$}
\hspace{\myx}
\hspace{-215pt}
}

%% file: editions/ten_zn.tex
\ensuremath{
\begin{matrix}
    \includegraphics[scale=1.7]{images/example_gral4leg.pdf}
\end{matrix}
\def \myx {40pt}
\def \myy {18pt}
\hspace{-\myx}
\raisebox{\myy}{$\dots$}
\hspace{\myx}
\def \myx {75pt}
\def \myy {-15pt}
\hspace{-\myx}
\raisebox{\myy}{$\dots$}
\hspace{\myx}
\def \myx {108pt}
\def \myy {-13pt}
\def \myangle {42}
\hspace{-\myx}
\raisebox{\myy}{
\rotatebox[origin=c]{\myangle}{$
\begin{tikzpicture}[line width=1.5pt, color=gray!165]
    \draw [-{Classical TikZ Rightarrow[length=3.5pt]}](0,0) -- (0.05,0);
\end{tikzpicture}
$}}
\hspace{\myx}
\def \myx {84.5pt}
\def \myy {-17pt}
\def \myangle {70}
\hspace{-\myx}
\raisebox{\myy}{
\rotatebox[origin=c]{\myangle}{$
\begin{tikzpicture}[line width=1.5pt, color=gray!165]
    \draw [-{Classical TikZ Rightarrow[length=3.5pt]}](0,0) -- (0.05,0);
\end{tikzpicture}
$}}
\hspace{\myx}
\def \myx {69pt}
\def \myy {3.7pt}
\def \myangle {11}
\hspace{-\myx}
\raisebox{\myy}{
\rotatebox[origin=c]{\myangle}{$
\begin{tikzpicture}[line width=1.5pt, color=gray!165]
    \draw [-{Classical TikZ Rightarrow[length=3.5pt]}](0,0) -- (0.05,0);
\end{tikzpicture}
$}}
\hspace{\myx}
\def \myx {117.5pt}
\def \myy {17pt}
\def \myangle {76}
\hspace{-\myx}
\raisebox{\myy}{
\rotatebox[origin=c]{\myangle}{$
\begin{tikzpicture}[line width=1.5pt, color=gray!165]
    \draw [-{Classical TikZ Rightarrow[length=3.5pt]}](0,0) -- (0.05,0);
\end{tikzpicture}
$}}
\hspace{\myx}
\def \myx {118pt}
\def \myy {0pt}
\hspace{-\myx}
\raisebox{\myy}{$\rrr{C}$}
\hspace{\myx}
\hspace{-58pt}
}

%% file: editions/Ztwo_ten.tex
\ensuremath{
\begin{matrix}
    \includegraphics[scale=1.1]{images/general_ten.pdf}
\end{matrix}
\def \myx {75pt}
\def \myy {15pt}
\hspace{-\myx}
\raisebox{\myy}{
$\ket{a_1{;}\alpha_1}$
}
\hspace{\myx}
\def \myx {100pt}
\def \myy {-17pt}
\hspace{-\myx}
\raisebox{\myy}{
$\ket{a_2{;}\alpha_2}$
}
\hspace{\myx}
\def \myx {95pt}
\def \myy {-14pt}
\hspace{-\myx}
\raisebox{\myy}{
$\ket{a_R{;}\alpha_R}$
}
\hspace{\myx}
\def \myx {150pt}
\def \myy {19pt}
\hspace{-\myx}
\raisebox{\myy}{
$C$
}
\hspace{\myx}
\def \myx {162pt}
\def \myy {0pt}
\hspace{-\myx}
\raisebox{\myy}{
${\dots}$
}
\hspace{\myx}
\def \myx {165pt}
\def \myy {15pt}
\hspace{-\myx}
\raisebox{\myy}{
$(a_\mathrm{tot})$
}
\hspace{\myx}
\def \mycompensate {110pt}
\hspace{-\mycompensate} 
\def \myx {118pt}
\def \myy {5pt}
\def \myangle {8}
\hspace{-\myx}
\raisebox{\myy}{
\rotatebox[origin=c]{\myangle}{$
\begin{tikzpicture}[line width=1.5pt, color=gray!165]
    \draw [-{Classical TikZ Rightarrow[length=3.5pt]}](0,0) -- (0.05,0);
\end{tikzpicture}
$}}
\hspace{\myx}
\def \myx {123pt}
\def \myy {-4pt}
\def \myangle {35}
\hspace{-\myx}
\raisebox{\myy}{
\rotatebox[origin=c]{\myangle}{$
\begin{tikzpicture}[line width=1.5pt, color=gray!165]
    \draw [-{Classical TikZ Rightarrow[length=3.5pt]}](0,0) -- (0.05,0);
\end{tikzpicture}
$}}
\hspace{\myx}
\def \myx {104pt}
\def \myy {-1.2pt}
\def \myangle {-25}
\hspace{-\myx}
\raisebox{\myy}{
\rotatebox[origin=c]{\myangle}{$
\begin{tikzpicture}[line width=1.5pt, color=gray!165]
    \draw [-{Classical TikZ Rightarrow[length=3.5pt]}](0,0) -- (0.05,0);
\end{tikzpicture}
$}}
\hspace{\myx}
\hspace{\mycompensate} 
\hspace{-190pt}
}

%% file: sections/E_ansatz_luca.tex
\section{Gauge invariant tensor networks}
\label{sec:ansatz}

Despite the fact that the projector onto a well-defined gauge-invariant sector $K(\rrr{\{g_i\}})$ spoils the tensor product structure of the Hilbert space, we can still write such a projector and, consequently, represent an arbitrary state of the gauge-invariant Hilbert space as a tensor network. Here, we follow the presentation in  \cite{tagliacozzo2014},  though equivalent formulations have been proposed in \cite{silvi2014,haegeman2015,zohar2016}. Recent advances include tensor networks for fermions at finite density interacting with Abelian  and non-Abelian gauge fields \cite{felser2020,robaina2021,cataldi2024}. For recent reviews on these topics see e.g. \cite{banuls2020,emonts2020,magnifico2024}.

We will focus on a pure gauge theory, although the extension to a gauge theory with matter is straightforward. In the pure gauge theory, the degrees of freedom live on the links $l$ of the lattice $\Lambda$.
To describe states living in a well-defined gauge sector $K(\rrr{\{g_i\}})$, we start by working in an enlarged Hilbert space composed of several irreducible representations $a_i$ of the Abelian group $G$, each with arbitrary multiplicity, ${\cal H}_i\simeq \oplus_i (a_i, \alpha_i)$. The group algebra is recovered by considering as the Hilbert space the direct sum of all possible irreducible representations, each appearing exactly once.

We write a state in the relevant gauge-invariant sector by projecting out, from the enlarged Hilbert space, those states that do not belong to $K(\{\rrr{g_i}\})$ \cite{buividovich2008}. The projection can be written as a Tensor Product Operator (see \cite{tagliacozzo2014}); alternatively, we can incorporate this projection into a variational ansatz. In particular, we can  copy the charge information  from the physical legs to the auxiliary legs and then use an appropriate tensor on each vertex $i$ to select the corresponding \rrr{subspace on the star with a certain eigenvalue of $A_i(h_1)$}, determined by the choice of the gauge sector $K(\rrr{\{g_i\}})$.

Copying the charge information from the physical to the auxiliary leg is achieved using so-called \textit{copy} tensors on the links, while the local constrain is imposed by the so-called \textit{vertex} tensors. In this way, all symmetry-breaking components are sent to zero, resulting in a final state that is locally symmetric, as defined in Eq.~\eqref{eq:gauge:K}.

\rrr{We consider the simple $\mathbb{Z}_2$ case to explicitly introduce both the TN and, later, the doubling strategy. The general \zn case will be explained afterwards.} Explicitly, following notation from Sec.~\ref{sec:TN}, and for a set of background charges \rrr{$\{g_i\}_{\forall i\in\mathcal{V}}$, where $g_i\in\{0,1\}$}, \ztwo locally symmetric states on an infinite \rrr{2-dimensional} square lattice are represented as
\begin{equation}
    \def \myshift {103pt}
    \ket\Psi=
    \hspace{-15pt}
    \input{editions/ansatz_layout}
\end{equation}
with vertex tensors $V_i$
\begin{gather}
    \input{editions/vertex_ten}
    =
    \begin{cases}
        \left[V_i\right]^{abcd}_{\alpha\beta\gamma\delta} & \text{if } a{+}b{+}c{+}d{=}g_i\modtwo
        \\
        0 & \text{otherwise}
    \end{cases}
    \def \myshift {100pt}
    \def \mybracelen {95pt}
    \hspace{-\myshift}
    \raisebox{15pt}{$\overbrace{\hspace{\mybracelen}}^\text{Gauss' law $\leftrightarrow$ \ztwo sum-rule}$},
    \label{eq:ansatz:vertex_ten}
\end{gather}
and copy tensors $C_l$
\begin{gather}
    \input{editions/copy_ten}
    =\begin{cases}
        \left[C_l\right]^{abc}_{\alpha 1\gamma} & \text{if }a=b=c \\
        0 & \text{otherwise}
    \end{cases}
    \def \myshift {43pt}
    \def \mybracelen {40pt}
    \hspace{-\myshift}
    \raisebox{15pt}{$\overbrace{\hspace{\mybracelen}}^\text{copy charge}$}
    .
    \label{eq:ansatz:Copyten}
\end{gather}

One can verify that the local constraint is satisfied at every vertex for such a TN. For example, consider a state where the local constraint is violated at a particular vertex, e.g., $0{+}1{+}1{+}1\modtwo{=}1$ with \rrr{conflicting} background charge $g_i{=}0$. In this case, the contraction of that vertex tensor would yield
\begin{equation}
    \input{editions/checkvertex1}
    \overset{(1)}{=}
    \sum_{\alpha\beta\gamma\delta}
    \input{editions/checkvertex2}
    \overset{(2)}{=}
    0.
    \label{eq:ansatz:checkvertex}
\end{equation}

The symmetry breaking component is successfully projected out by the ansatz. Step $(1)$ copies the physical charge and sends it to the vertex, while step $(2)$ results from $V_i$ evaluating to zero for each term in the summation, since the constrain of Eq.~\eqref{eq:ansatz:vertex_ten} is never met. This occurs for every component on every vertex, ensuring that every state $\ket\Psi$ represented by the ansatz is locally symmetric\footnote{It can also be shown that every locally symmetric state can be expressed by this ansatz.}.

While the tensors $V_i$ are \ztwo  symmetric tensors, the tensors $C_l$ are not, as can be verified by direct inspection. As a result, using this  formulation for  gauge invariant sates, the tensor network contains a mixture of symmetric tensors and sparse but non-symmetric tensors.  Therefore, we cannot directly  exploit the sparse structure of the vertex tensors.
In Sec.~\ref{sec:doublesym}, we will provide an alternative formulation that relies solely on constituent tensors that are symmetric. This new formulation allows us to exploit the sparse nature of the elementary tensors, speeding up the calculations while explicitly preserving gauge invariance and, if necessary, targeting a specific gauge sector. Before doing this, we review some of the theoretical consequences of obtaining a gauge invariant formulation of tensor networks. Such tensor networks indeed provide explicit examples of i) duality transformations between spin and gauge systems, ii)  of the use of Clifford gates to reduced entanglement before using full variational tensor networks, and iii) of finite depth circuits that are able to prepare topological states. These theoretical implications  justify the presentation of the ansatz separately from its numerical implementation. We thus review these three implications here for completeness. The reader interested to the new ansatz solely can skip the next subsection.

\subsection{Theoretical implications of gauge invariant tensor networks}

One of the key properties of the gauge invariant formulation of the PEPS tensor network  is that it can be implemented as a finite depth circuit. Such a finite depth circuit allows to prepare a topologically ordered state starting from a product state. Indeed the Toric code \rrr{(TC)} ground state \cite{kitaev2003} is the ground state of the \ztwo LGT in the limit of weak coupling $g_\rrr{KS} \to 0$. In that limit the ground state is just the equal superposition of all states in the $K(0,\dots,0)$ gauge sector.

Such a ground state can be obtained (up to normalization) by the the PEPS TN anstaz, where the vertex tensors $V_i$ contains ones in all possible symmetric entries.

We now review how such an ansatz is equivalent to a finite depth unitary circuit plus measurements. In the case we are discussing we can use elementary two body gates taken from the Clifford group.

\begin{figure}
 \includegraphics[width=0.7\textwidth]{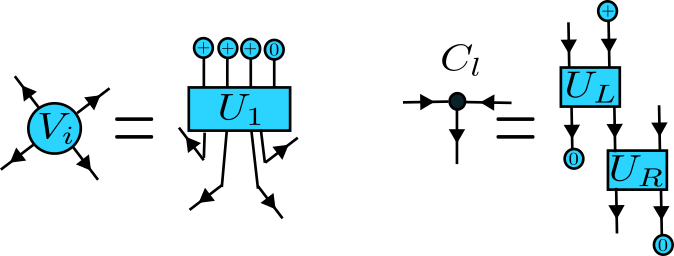}
 \caption{The equivalence of the gauge invariant ansatz presented in \cite{tagliacozzo2014} with a finite depth unitary circuit followed by measurements. The vertex tensor are obtained by acting on four qubits on a  specific product state, \rrr{and then two adjacent vertex tensor are connected via two extra unitary gates, that act on two outgoing legs of the vertex circuit plus an extra qubit initialized in a product state. Of the three outgoing qubit per lattice link, two needs to be projected on the state $\ket{0}$} in order to recover the desired ground state of the Toric code, a paradigmatic example of a state with \ztwo topological order.\label{fig:unitary}}
\end{figure}

Such construction is summarized in Fig.~\ref{fig:unitary}. There, we show that the vertex tensor can be obtain by acting with a unitary gate, labeled as $U_1$ in the figure, on four qubits appropriately initialized in a product state. For the specific form of $U_1$  in terms of elementary C-Not gates the choice of the  initial state should be $\ket{+++0}$. The first three $\ket{+}$ ensure we are considering an equal superposition of all gauge invariant states, while the last $\ket{0}$ selects the relevant gauge invariant sector. Each outgoing leg of the vertex tensor then needs to be concatenated with two extra unitaries, called  $U_L, U_R$ in the figure, that implement the copy tensors.  $U_L, U_R$  can again be expressed in terms of C-Not gates, as specified below, and as a consequence they need to act on the state $\ket{+}$ again to ensure that we are considering a linear superposition of all possible gauge invariant state. If we are interesting in building a projector onto all possible states of $K(0,\dots, 0)$, such a state (encircled in the figure) can be removed. Finally, two of the outgoing legs of $U_L$ and $U_R$ need to be projected onto the state $\ket{0}$ in order to recover the desired Toric code ground state or the projector onto the sector $K(0,\dots, 0)$ of the \ztwo LTG.
It is important to notice, that by selecting different initial state on which $U_1$ acts (e.g. $\ket{+++,1}$), we can build similarly the projector onto any gauge invariant sector $K(1,0,\dots,0)$.

The choice of the product state we use at the beginning of the unitary circuit, and at its end, depend on the specific form of $U_1$ and $U_L$ and $U_R$. In the case we are discussing we have used elementary two-body gates built out from the C-Not gate
\begin{equation}
 U_c =\ket{0}\bra{0}\otimes\mathbb{I} + \ket{1}\bra{1}\otimes\sigma_x. \label{eq:cnot}
\end{equation}

\begin{figure}
 \includegraphics[width=0.7\textwidth]{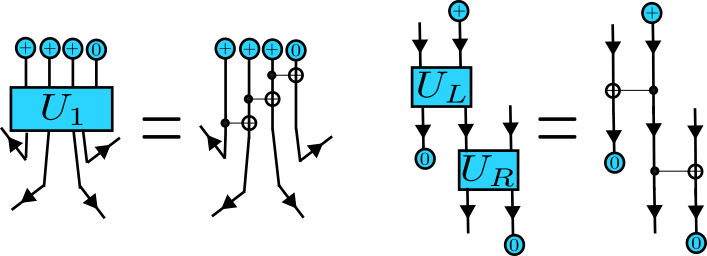}
\caption{The specific form of $U_1$, $U_L$ and $U_R$ in terms of C-Not gates, depicted by their standard quantum circuit encoding. \label{fig:cnots}}
\end{figure}

The specific form of $U_1$ and $U_L$, $U_R$ is depicted in Fig.~\ref{fig:cnots}

Given the fact that the uniarty circuit maps the TC ground state into a product state and is completely based on Clifford gates, the same  gauge invariant tensor network ansatz can also be interpreted as a Clifford circuit that disentangle the TC ground state.  Therefore, such a TPO projector (and its predecessor MERA TN, aslo based on the concatenation of Clifford gates and variationally optimized tensors discussed in \cite{tagliacozzo2011}) are examples of what today one would call Clifford enhanced TN \cite{fux2024,mello2024}. 

Finally, one can also interpret the projector onto the gauge invariant Hilbert space as the generator of a duality transformation that acts  between the LGT and a non-symmetric theory. Such an interpretation was explicitly discussed in 
\cite{tagliacozzo2011}, where it was explained how the MERA structure of the network allowed to obtain a linear depth circuit that would implement a specific version of the duality transformation that would map local operators into local operators. 

The PEPS structure, on the other hand include a finite depth circuit followed by measurments. Such measurements can in principle spoint the locality of the causal cone of the network and thus one needs to relax the idea that local opeartors are sent to local operators and accept that local operators, in general are sent into TPO operators. The TN TPO projector  thus provides an alternative construction of a duality transformation along the lines discussed originally in \cite{tagliacozzo2011} and more recently in  \cite{tagliacozzo2014}.


%% file: editions/ansatz_layout.tex
\ensuremath{
\begin{matrix}
    \includegraphics[scale=0.7]{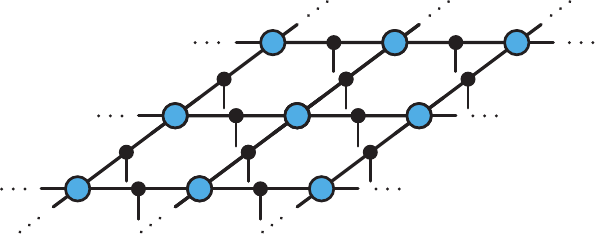}
\end{matrix}
\def \myx {110pt}
\def \myy {13pt}
\hspace{-\myx}
\raisebox{\myy}{
$V_i$
}
\hspace{\myx}
\def \myx {95pt}
\def \myy {10pt}
\hspace{-\myx}
\raisebox{\myy}{
$C_l$
}
\hspace{\myx}
\def \myx {137pt}
\def \myy {-7pt}
\hspace{-\myx}
\raisebox{\myy}{
$g_i$
}
\hspace{\myx}
\def \myx {146pt}
\def \myy {25.85pt}
\hspace{-\myx}
\raisebox{\myy}{$
\begin{tikzpicture}[line width=1.5pt, color=gray!165]
    \draw [-{Classical TikZ Rightarrow[length=3.5pt]}](0,0) -- (0.05,0);
\end{tikzpicture}
$}
\hspace{\myx}
\def \myx {132pt}
\def \myy {25.85pt}
\hspace{-\myx}
\raisebox{\myy}{$
\begin{tikzpicture}[line width=1.5pt, color=gray!165]
    \draw [-{Classical TikZ Rightarrow[length=3.5pt]}](0,0) -- (0.05,0);
\end{tikzpicture}
$}
\hspace{\myx}
\def \myx {113.5pt}
\def \myy {25.85pt}
\hspace{-\myx}
\raisebox{\myy}{$
\begin{tikzpicture}[line width=1.5pt, color=gray!165]
    \draw [-{Classical TikZ Rightarrow[length=3.5pt]}](0,0) -- (0.05,0);
\end{tikzpicture}
$}
\hspace{\myx}
\def \myx {100pt}
\def \myy {25.85pt}
\hspace{-\myx}
\raisebox{\myy}{$
\begin{tikzpicture}[line width=1.5pt, color=gray!165]
    \draw [-{Classical TikZ Rightarrow[length=3.5pt]}](0,0) -- (0.05,0);
\end{tikzpicture}
$}
\hspace{\myx}
\def \myh {1.15pt}
\def \mycompensatex {50pt}
\hspace{-\mycompensatex}
\def \myx {146pt}
\def \myy {\myh}
\hspace{-\myx}
\raisebox{\myy}{$
\begin{tikzpicture}[line width=1.5pt, color=gray!165]
    \draw [-{Classical TikZ Rightarrow[length=3.5pt]}](0,0) -- (0.05,0);
\end{tikzpicture}
$}
\hspace{\myx}
\def \myx {132pt}
\def \myy {\myh}
\hspace{-\myx}
\raisebox{\myy}{$
\begin{tikzpicture}[line width=1.5pt, color=gray!165]
    \draw [-{Classical TikZ Rightarrow[length=3.5pt]}](0,0) -- (0.05,0);
\end{tikzpicture}
$}
\hspace{\myx}
\def \myx {113.5pt}
\def \myy {\myh}
\hspace{-\myx}
\raisebox{\myy}{$
\begin{tikzpicture}[line width=1.5pt, color=gray!165]
    \draw [-{Classical TikZ Rightarrow[length=3.5pt]}](0,0) -- (0.05,0);
\end{tikzpicture}
$}
\hspace{\myx}
\def \myx {100pt}
\def \myy {\myh}
\hspace{-\myx}
\raisebox{\myy}{$
\begin{tikzpicture}[line width=1.5pt, color=gray!165]
    \draw [-{Classical TikZ Rightarrow[length=3.5pt]}](0,0) -- (0.05,0);
\end{tikzpicture}
$}
\hspace{\myx}
\hspace{\mycompensatex}
\def \myh {-23.4pt}
\def \mycompensatex {100.5pt}
\hspace{-\mycompensatex}
\def \myx {146pt}
\def \myy {\myh}
\hspace{-\myx}
\raisebox{\myy}{$
\begin{tikzpicture}[line width=1.5pt, color=gray!165]
    \draw [-{Classical TikZ Rightarrow[length=3.5pt]}](0,0) -- (0.05,0);
\end{tikzpicture}
$}
\hspace{\myx}
\def \myx {132pt}
\def \myy {\myh}
\hspace{-\myx}
\raisebox{\myy}{$
\begin{tikzpicture}[line width=1.5pt, color=gray!165]
    \draw [-{Classical TikZ Rightarrow[length=3.5pt]}](0,0) -- (0.05,0);
\end{tikzpicture}
$}
\hspace{\myx}
\def \myx {113.5pt}
\def \myy {\myh}
\hspace{-\myx}
\raisebox{\myy}{$
\begin{tikzpicture}[line width=1.5pt, color=gray!165]
    \draw [-{Classical TikZ Rightarrow[length=3.5pt]}](0,0) -- (0.05,0);
\end{tikzpicture}
$}
\hspace{\myx}
\def \myx {100pt}
\def \myy {\myh}
\hspace{-\myx}
\raisebox{\myy}{$
\begin{tikzpicture}[line width=1.5pt, color=gray!165]
    \draw [-{Classical TikZ Rightarrow[length=3.5pt]}](0,0) -- (0.05,0);
\end{tikzpicture}
$}
\hspace{\myx}
\hspace{\mycompensatex}
\def \myx {223pt}
\def \myy {19.1pt}
\hspace{-\myx}
\raisebox{\myy}{
\rotatebox[origin=c]{218}{$
\begin{tikzpicture}[line width=1.5pt, color=gray!165]
    \draw [-{Classical TikZ Rightarrow[length=3.5pt]}](0,0) -- (0.05,0);
\end{tikzpicture}
$}}
\hspace{\myx}
\def \myx {248.2pt}
\def \myy {8.3pt}
\hspace{-\myx}
\raisebox{\myy}{
\rotatebox[origin=c]{218}{$
\begin{tikzpicture}[line width=1.5pt, color=gray!165]
    \draw [-{Classical TikZ Rightarrow[length=3.5pt]}](0,0) -- (0.05,0);
\end{tikzpicture}
$}}
\hspace{\myx}
\def \myx {277pt}
\def \myy {-5pt}
\hspace{-\myx}
\raisebox{\myy}{
\rotatebox[origin=c]{218}{$
\begin{tikzpicture}[line width=1.5pt, color=gray!165]
    \draw [-{Classical TikZ Rightarrow[length=3.5pt]}](0,0) -- (0.05,0);
\end{tikzpicture}
$}}
\hspace{\myx}
\def \myx {302.5pt}
\def \myy {-16pt}
\hspace{-\myx}
\raisebox{\myy}{
\rotatebox[origin=c]{218}{$
\begin{tikzpicture}[line width=1.5pt, color=gray!165]
    \draw [-{Classical TikZ Rightarrow[length=3.5pt]}](0,0) -- (0.05,0);
\end{tikzpicture}
$}}
\hspace{\myx}
\def \mycompensate {2pt}
\hspace{-\mycompensate}
\def \myx {223pt}
\def \myy {19.1pt}
\hspace{-\myx}
\raisebox{\myy}{
\rotatebox[origin=c]{218}{$
\begin{tikzpicture}[line width=1.5pt, color=gray!165]
    \draw [-{Classical TikZ Rightarrow[length=3.5pt]}](0,0) -- (0.05,0);
\end{tikzpicture}
$}}
\hspace{\myx}
\def \myx {248.2pt}
\def \myy {8.3pt}
\hspace{-\myx}
\raisebox{\myy}{
\rotatebox[origin=c]{218}{$
\begin{tikzpicture}[line width=1.5pt, color=gray!165]
    \draw [-{Classical TikZ Rightarrow[length=3.5pt]}](0,0) -- (0.05,0);
\end{tikzpicture}
$}}
\hspace{\myx}
\def \myx {277pt}
\def \myy {-5pt}
\hspace{-\myx}
\raisebox{\myy}{
\rotatebox[origin=c]{218}{$
\begin{tikzpicture}[line width=1.5pt, color=gray!165]
    \draw [-{Classical TikZ Rightarrow[length=3.5pt]}](0,0) -- (0.05,0);
\end{tikzpicture}
$}}
\hspace{\myx}
\def \myx {302.5pt}
\def \myy {-16pt}
\hspace{-\myx}
\raisebox{\myy}{
\rotatebox[origin=c]{218}{$
\begin{tikzpicture}[line width=1.5pt, color=gray!165]
    \draw [-{Classical TikZ Rightarrow[length=3.5pt]}](0,0) -- (0.05,0);
\end{tikzpicture}
$}}
\hspace{\myx}
\hspace{\mycompensate}
\def \mycompensate {4.3pt}
\hspace{-\mycompensate}
\def \myx {223pt}
\def \myy {19.1pt}
\hspace{-\myx}
\raisebox{\myy}{
\rotatebox[origin=c]{218}{$
\begin{tikzpicture}[line width=1.5pt, color=gray!165]
    \draw [-{Classical TikZ Rightarrow[length=3.5pt]}](0,0) -- (0.05,0);
\end{tikzpicture}
$}}
\hspace{\myx}
\def \myx {248.2pt}
\def \myy {8.3pt}
\hspace{-\myx}
\raisebox{\myy}{
\rotatebox[origin=c]{218}{$
\begin{tikzpicture}[line width=1.5pt, color=gray!165]
    \draw [-{Classical TikZ Rightarrow[length=3.5pt]}](0,0) -- (0.05,0);
\end{tikzpicture}
$}}
\hspace{\myx}
\def \myx {277pt}
\def \myy {-5pt}
\hspace{-\myx}
\raisebox{\myy}{
\rotatebox[origin=c]{218}{$
\begin{tikzpicture}[line width=1.5pt, color=gray!165]
    \draw [-{Classical TikZ Rightarrow[length=3.5pt]}](0,0) -- (0.05,0);
\end{tikzpicture}
$}}
\hspace{\myx}
\def \myx {302.5pt}
\def \myy {-16pt}
\hspace{-\myx}
\raisebox{\myy}{
\rotatebox[origin=c]{218}{$
\begin{tikzpicture}[line width=1.5pt, color=gray!165]
    \draw [-{Classical TikZ Rightarrow[length=3.5pt]}](0,0) -- (0.05,0);
\end{tikzpicture}
$}}
\hspace{\myx}
\hspace{\mycompensate}
\def \myx {360.05pt}
\def \myy {7.2pt}
\hspace{-\myx}
\raisebox{\myy}{
\rotatebox[origin=c]{90}{$
\begin{tikzpicture}[line width=1.5pt, color=gray!165]
    \draw [-{Classical TikZ Rightarrow[length=3.5pt]}](0,0) -- (0.05,0);
\end{tikzpicture}
$}}
\hspace{\myx}
\def \myx {403pt}
\def \myy {-17pt}
\hspace{-\myx}
\raisebox{\myy}{
\rotatebox[origin=c]{90}{$
\begin{tikzpicture}[line width=1.5pt, color=gray!165]
    \draw [-{Classical TikZ Rightarrow[length=3.5pt]}](0,0) -- (0.05,0);
\end{tikzpicture}
$}}
\hspace{\myx}
\def \mycompensate {-21pt}
\hspace{-\mycompensate}
\def \myx {360.05pt}
\def \myy {7.2pt}
\hspace{-\myx}
\raisebox{\myy}{
\rotatebox[origin=c]{90}{$
\begin{tikzpicture}[line width=1.5pt, color=gray!165]
    \draw [-{Classical TikZ Rightarrow[length=3.5pt]}](0,0) -- (0.05,0);
\end{tikzpicture}
$}}
\hspace{\myx}
\def \myx {403pt}
\def \myy {-17pt}
\hspace{-\myx}
\raisebox{\myy}{
\rotatebox[origin=c]{90}{$
\begin{tikzpicture}[line width=1.5pt, color=gray!165]
    \draw [-{Classical TikZ Rightarrow[length=3.5pt]}](0,0) -- (0.05,0);
\end{tikzpicture}
$}}
\hspace{\myx}
\hspace{\mycompensate}
\def \mycompensate {-42pt}
\hspace{-\mycompensate}
\def \myx {360.05pt}
\def \myy {7.2pt}
\hspace{-\myx}
\raisebox{\myy}{
\rotatebox[origin=c]{90}{$
\begin{tikzpicture}[line width=1.5pt, color=gray!165]
    \draw [-{Classical TikZ Rightarrow[length=3.5pt]}](0,0) -- (0.05,0);
\end{tikzpicture}
$}}
\hspace{\myx}
\def \myx {403pt}
\def \myy {-17pt}
\hspace{-\myx}
\raisebox{\myy}{
\rotatebox[origin=c]{90}{$
\begin{tikzpicture}[line width=1.5pt, color=gray!165]
    \draw [-{Classical TikZ Rightarrow[length=3.5pt]}](0,0) -- (0.05,0);
\end{tikzpicture}
$}}
\hspace{\myx}
\hspace{\mycompensate}
\def \myx {383.4pt}
\def \myy {20pt}
\hspace{-\myx}
\raisebox{\myy}{
\rotatebox[origin=c]{90}{$
\begin{tikzpicture}[line width=1.5pt, color=gray!165]
    \draw [-{Classical TikZ Rightarrow[length=3.5pt]}](0,0) -- (0.05,0);
\end{tikzpicture}
$}}
\hspace{\myx}
\def \myx {426.4pt}
\def \myy {-5.5pt}
\hspace{-\myx}
\raisebox{\myy}{
\rotatebox[origin=c]{90}{$
\begin{tikzpicture}[line width=1.5pt, color=gray!165]
    \draw [-{Classical TikZ Rightarrow[length=3.5pt]}](0,0) -- (0.05,0);
\end{tikzpicture}
$}}
\hspace{\myx}
\def \myx {469.4pt}
\def \myy {-30pt}
\hspace{-\myx}
\raisebox{\myy}{
\rotatebox[origin=c]{90}{$
\begin{tikzpicture}[line width=1.5pt, color=gray!165]
    \draw [-{Classical TikZ Rightarrow[length=3.5pt]}](0,0) -- (0.05,0);
\end{tikzpicture}
$}}
\hspace{\myx}
\def \mycompensate {-10.9pt}
\hspace{-\mycompensate}
\def \myx {383.4pt}
\def \myy {20pt}
\hspace{-\myx}
\raisebox{\myy}{
\rotatebox[origin=c]{90}{$
\begin{tikzpicture}[line width=1.5pt, color=gray!165]
    \draw [-{Classical TikZ Rightarrow[length=3.5pt]}](0,0) -- (0.05,0);
\end{tikzpicture}
$}}
\hspace{\myx}
\def \myx {426.4pt}
\def \myy {-5.5pt}
\hspace{-\myx}
\raisebox{\myy}{
\rotatebox[origin=c]{90}{$
\begin{tikzpicture}[line width=1.5pt, color=gray!165]
    \draw [-{Classical TikZ Rightarrow[length=3.5pt]}](0,0) -- (0.05,0);
\end{tikzpicture}
$}}
\hspace{\myx}
\def \myx {469.4pt}
\def \myy {-30pt}
\hspace{-\myx}
\raisebox{\myy}{
\rotatebox[origin=c]{90}{$
\begin{tikzpicture}[line width=1.5pt, color=gray!165]
    \draw [-{Classical TikZ Rightarrow[length=3.5pt]}](0,0) -- (0.05,0);
\end{tikzpicture}
$}}
\hspace{\myx}
\hspace{\mycompensate}
\hspace{-350pt},
}

%% file: editions/vertex_ten.tex
\ensuremath{
\begin{matrix}
    \includegraphics[scale=1.1]{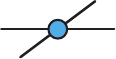}
\end{matrix}
\def \myx {25pt}
\def \myy {25pt}
\hspace{-\myx}
\raisebox{\myy}{
$\ket{a{;}\alpha}$
}
\hspace{\myx}
\def \myx {90pt}
\def \myy {-22pt}
\hspace{-\myx}
\raisebox{\myy}{
$\ket{c{;}\gamma}$
}
\hspace{\myx}
\def \myx {75pt}
\def \myy {-7pt}
\hspace{-\myx}
\raisebox{\myy}{
$\ket{b{;}\beta}$
}
\hspace{\myx}
\def \myx {145pt}
\def \myy {10pt}
\hspace{-\myx}
\raisebox{\myy}{
$\ket{d{;}\delta}$
}
\hspace{\myx}
\def \myx {145pt}
\def \myy {15pt}
\hspace{-\myx}
\raisebox{\myy}{
$V_i$
}
\hspace{\myx}
\def \myx {160pt}
\def \myy {-10pt}
\hspace{-\myx}
\raisebox{\myy}{
$g_i$
}
\hspace{\myx}
\def \mycompensate {-20pt}
\hspace{-\mycompensate}
\def \myx {169pt}
\def \myy {1.0pt}
\hspace{-\myx}
\raisebox{\myy}{$
\begin{tikzpicture}[line width=1.5pt, color=gray!165]
    \draw [-{Classical TikZ Rightarrow[length=3.5pt]}](0,0) -- (0.05,0);
\end{tikzpicture}
$}
\hspace{\myx}
\def \myx {214pt}
\def \myy {1.0pt}
\hspace{-\myx}
\raisebox{\myy}{$
\begin{tikzpicture}[line width=1.5pt, color=gray!165]
    \draw [-{Classical TikZ Rightarrow[length=3.5pt]}](0,0) -- (0.05,0);
\end{tikzpicture}
$}
\hspace{\myx}
\def \myx {189.5pt}
\def \myy {11.5pt}
\hspace{-\myx}
\raisebox{\myy}{
\rotatebox[origin=c]{218}{$
\begin{tikzpicture}[line width=1.5pt, color=gray!165]
    \draw [-{Classical TikZ Rightarrow[length=3.5pt]}](0,0) -- (0.05,0);
\end{tikzpicture}
$}}
\hspace{\myx}
\def \myx {228.5pt}
\def \myy {-9.5pt}
\hspace{-\myx}
\raisebox{\myy}{
\rotatebox[origin=c]{218}{$
\begin{tikzpicture}[line width=1.5pt, color=gray!165]
    \draw [-{Classical TikZ Rightarrow[length=3.5pt]}](0,0) -- (0.05,0);
\end{tikzpicture}
$}}
\hspace{\myx}
\hspace{\mycompensate}
\hspace{-165pt}
}

%% file: editions/copy_ten.tex
\ensuremath{
\begin{matrix}
    \includegraphics[scale=1.2]{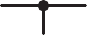}
\end{matrix}
\def \myx {77pt}
\def \myy {9pt}
\hspace{-\myx}
\raisebox{\myy}{
$\ket{a{;}\alpha}$
}
\hspace{\myx}
\def \myx {28pt}
\def \myy {9pt}
\hspace{-\myx}
\raisebox{\myy}{
$\ket{c{;}\gamma}$
}
\hspace{\myx}
\def \myx {91pt}
\def \myy {-16pt}
\hspace{-\myx}
\raisebox{\myy}{
$\ket{b{;}1}$
}
\hspace{\myx}
\def \myx {112pt}
\def \myy {18pt}
\hspace{-\myx}
\raisebox{\myy}{
$C_l$
}
\hspace{\myx}
\hspace{15.5pt}
\def \myx {125pt}
\def \myy {7.5pt}
\hspace{-\myx}
\raisebox{\myy}{$
\begin{tikzpicture}[line width=1.5pt, color=gray!165]
    \draw [-{Classical TikZ Rightarrow[length=3.5pt]}](0,0) -- (0.05,0);
\end{tikzpicture}
$}
\hspace{\myx}
\def \myx {156pt}
\def \myy {7.5pt}
\hspace{-\myx}
\raisebox{\myy}{$
\begin{tikzpicture}[line width=1.5pt, color=gray!165]
    \draw [-{Classical TikZ Rightarrow[length=3.5pt]}](0,0) -- (0.05,0);
\end{tikzpicture}
$}
\hspace{\myx}
\def \myx {152.1pt}
\def \myy {-2pt}
\hspace{-\myx}
\raisebox{\myy}{
\rotatebox[origin=c]{90}{$
\begin{tikzpicture}[line width=1.5pt, color=gray!165]
    \draw [-{Classical TikZ Rightarrow[length=3.5pt]}](0,0) -- (0.05,0);
\end{tikzpicture}
$}}
\hspace{\myx}
\hspace{-105pt}
}

%% file: editions/checkvertex1.tex
\ensuremath{
\begin{matrix}
    \includegraphics[scale=1.05]{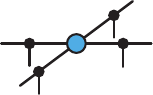}
\end{matrix}
\def \myx {52pt}
\def \myy {-10pt}
\hspace{-\myx}
\raisebox{\myy}{
$g_i{=}0$
}
\hspace{\myx}
\hspace{-3pt}
\def \myx {115pt}
\def \myy {-5pt}
\hspace{-\myx}
\raisebox{\myy}{
$\ket{1{;}1}$
}
\hspace{\myx}
\def \myx {120pt}
\def \myy {-30pt}
\hspace{-\myx}
\raisebox{\myy}{
$\ket{1{;}1}$
}
\hspace{\myx}
\def \myx {105pt}
\def \myy {-15pt}
\hspace{-\myx}
\raisebox{\myy}{
$\ket{1{;}1}$
}
\hspace{\myx}
\def \myx {120pt}
\def \myy {15pt}
\hspace{-\myx}
\raisebox{\myy}{
$\ket{0{;}1}$
}
\hspace{\myx}
\def \myx {177pt}
\def \myy {17pt}
\hspace{-\myx}
\raisebox{\myy}{
$V_i$
}
\hspace{\myx}
\def \myarrowlen {3.5pt}
\def \myx {172pt}
\def \myy {2.8pt}
\hspace{-\myx}
\raisebox{\myy}{$
\begin{tikzpicture}[line width=1.5pt, color=gray!165]
    \draw [-{Classical TikZ Rightarrow[length=\myarrowlen]}](0,0) -- (0.05,0);
\end{tikzpicture}
$}
\hspace{\myx}
\def \myx {202pt}
\def \myy {2.8pt}
\hspace{-\myx}
\raisebox{\myy}{$
\begin{tikzpicture}[line width=1.5pt, color=gray!165]
    \draw [-{Classical TikZ Rightarrow[length=\myarrowlen]}](0,0) -- (0.05,0);
\end{tikzpicture}
$}
\hspace{\myx}
\def \myx {188pt}
\def \myy {11pt}
\hspace{-\myx}
\raisebox{\myy}{
\rotatebox[origin=c]{218}{$
\begin{tikzpicture}[line width=1.5pt, color=gray!165]
    \draw [-{Classical TikZ Rightarrow[length=\myarrowlen]}](0,0) -- (0.05,0);
\end{tikzpicture}
$}}
\hspace{\myx}
\def \myx {220.5pt}
\def \myy {-5pt}
\hspace{-\myx}
\raisebox{\myy}{
\rotatebox[origin=c]{218}{$
\begin{tikzpicture}[line width=1.5pt, color=gray!165]
    \draw [-{Classical TikZ Rightarrow[length=\myarrowlen]}](0,0) -- (0.05,0);
\end{tikzpicture}
$}}
\hspace{\myx}
\def \myx {183pt}
\def \myy {2.8pt}
\hspace{-\myx}
\raisebox{\myy}{$
\begin{tikzpicture}[line width=1.5pt, color=gray!165]
    \draw [-{Classical TikZ Rightarrow[length=\myarrowlen]}](0,0) -- (0.05,0);
\end{tikzpicture}
$}
\hspace{\myx}
\def \myx {251pt}
\def \myy {2.8pt}
\hspace{-\myx}
\raisebox{\myy}{$
\begin{tikzpicture}[line width=1.5pt, color=gray!165]
    \draw [-{Classical TikZ Rightarrow[length=\myarrowlen]}](0,0) -- (0.05,0);
\end{tikzpicture}
$}
\hspace{\myx}
\def \myx {204pt}
\def \myy {21.8pt}
\hspace{-\myx}
\raisebox{\myy}{
\rotatebox[origin=c]{218}{$
\begin{tikzpicture}[line width=1.5pt, color=gray!165]
    \draw [-{Classical TikZ Rightarrow[length=\myarrowlen]}](0,0) -- (0.05,0);
\end{tikzpicture}
$}}
\hspace{\myx}
\def \myx {264.5pt}
\def \myy {-15.5pt}
\hspace{-\myx}
\raisebox{\myy}{
\rotatebox[origin=c]{218}{$
\begin{tikzpicture}[line width=1.5pt, color=gray!165]
    \draw [-{Classical TikZ Rightarrow[length=\myarrowlen]}](0,0) -- (0.05,0);
\end{tikzpicture}
$}}
\hspace{\myx}
\def \myx {226.8pt}
\def \myy {-4pt}
\hspace{-\myx}
\raisebox{\myy}{
\rotatebox[origin=c]{90}{$
\begin{tikzpicture}[line width=1.5pt, color=gray!165]
    \draw [-{Classical TikZ Rightarrow[length=\myarrowlen]}](0,0) -- (0.05,0);
\end{tikzpicture}
$}}
\hspace{\myx}
\def \myx {284.3pt}
\def \myy {-4pt}
\hspace{-\myx}
\raisebox{\myy}{
\rotatebox[origin=c]{90}{$
\begin{tikzpicture}[line width=1.5pt, color=gray!165]
    \draw [-{Classical TikZ Rightarrow[length=\myarrowlen]}](0,0) -- (0.05,0);
\end{tikzpicture}
$}}
\hspace{\myx}
\def \myx {251.5pt}
\def \myy {10pt}
\hspace{-\myx}
\raisebox{\myy}{
\rotatebox[origin=c]{90}{$
\begin{tikzpicture}[line width=1.5pt, color=gray!165]
    \draw [-{Classical TikZ Rightarrow[length=\myarrowlen]}](0,0) -- (0.05,0);
\end{tikzpicture}
$}}
\hspace{\myx}
\def \myx {299.5pt}
\def \myy {-18.8pt}
\hspace{-\myx}
\raisebox{\myy}{
\rotatebox[origin=c]{90}{$
\begin{tikzpicture}[line width=1.5pt, color=gray!165]
    \draw [-{Classical TikZ Rightarrow[length=\myarrowlen]}](0,0) -- (0.05,0);
\end{tikzpicture}
$}}
\hspace{\myx}
\hspace{-240pt}
}

%% file: editions/checkvertex2.tex
\ensuremath{
\begin{matrix}
    \includegraphics[scale=1.05]{images/vertex_neighbours.pdf}
\end{matrix}
\def \myx {67pt}
\def \myy {13pt}
\hspace{-\myx}
\raisebox{\myy}{
$\ket{1{;}\delta}$
}
\hspace{\myx}
\def \myx {80pt}
\def \myy {-19pt}
\hspace{-\myx}
\raisebox{\myy}{
$\ket{1{;}\gamma}$
}
\hspace{\myx}
\def \myx {95pt}
\def \myy {-8pt}
\hspace{-\myx}
\raisebox{\myy}{
$\ket{1{;}\beta}$
}
\hspace{\myx}
\def \myx {131pt}
\def \myy {24pt}
\hspace{-\myx}
\raisebox{\myy}{
$\ket{0{;}\alpha}$
}
\hspace{\myx}
\def \myx {175pt}
\def \myy {29pt}
\hspace{-\myx}
\raisebox{\myy}{
$V_i$
}
\hspace{\myx}
\def \myx {155pt}
\def \myy {-20pt}
\hspace{-\myx}
\raisebox{\myy}{
$g_i{=}0$
}
\hspace{\myx}
\hspace{-3pt}
\def \myarrowlen {3.5pt}
\def\mycompensate {3.2pt}
\hspace{-\mycompensate}
\def \myx {172pt}
\def \myy {2.8pt}
\hspace{-\myx}
\raisebox{\myy}{$
\begin{tikzpicture}[line width=1.5pt, color=gray!165]
    \draw [-{Classical TikZ Rightarrow[length=\myarrowlen]}](0,0) -- (0.05,0);
\end{tikzpicture}
$}
\hspace{\myx}
\def \myx {202pt}
\def \myy {2.8pt}
\hspace{-\myx}
\raisebox{\myy}{$
\begin{tikzpicture}[line width=1.5pt, color=gray!165]
    \draw [-{Classical TikZ Rightarrow[length=\myarrowlen]}](0,0) -- (0.05,0);
\end{tikzpicture}
$}
\hspace{\myx}
\def \myx {188pt}
\def \myy {11pt}
\hspace{-\myx}
\raisebox{\myy}{
\rotatebox[origin=c]{218}{$
\begin{tikzpicture}[line width=1.5pt, color=gray!165]
    \draw [-{Classical TikZ Rightarrow[length=\myarrowlen]}](0,0) -- (0.05,0);
\end{tikzpicture}
$}}
\hspace{\myx}
\def \myx {220.5pt}
\def \myy {-5pt}
\hspace{-\myx}
\raisebox{\myy}{
\rotatebox[origin=c]{218}{$
\begin{tikzpicture}[line width=1.5pt, color=gray!165]
    \draw [-{Classical TikZ Rightarrow[length=\myarrowlen]}](0,0) -- (0.05,0);
\end{tikzpicture}
$}}
\hspace{\myx}
\def \myx {183pt}
\def \myy {2.8pt}
\hspace{-\myx}
\raisebox{\myy}{$
\begin{tikzpicture}[line width=1.5pt, color=gray!165]
    \draw [-{Classical TikZ Rightarrow[length=\myarrowlen]}](0,0) -- (0.05,0);
\end{tikzpicture}
$}
\hspace{\myx}
\def \myx {251pt}
\def \myy {2.8pt}
\hspace{-\myx}
\raisebox{\myy}{$
\begin{tikzpicture}[line width=1.5pt, color=gray!165]
    \draw [-{Classical TikZ Rightarrow[length=\myarrowlen]}](0,0) -- (0.05,0);
\end{tikzpicture}
$}
\hspace{\myx}
\def \myx {204pt}
\def \myy {21.8pt}
\hspace{-\myx}
\raisebox{\myy}{
\rotatebox[origin=c]{218}{$
\begin{tikzpicture}[line width=1.5pt, color=gray!165]
    \draw [-{Classical TikZ Rightarrow[length=\myarrowlen]}](0,0) -- (0.05,0);
\end{tikzpicture}
$}}
\hspace{\myx}
\def \myx {264.5pt}
\def \myy {-15.5pt}
\hspace{-\myx}
\raisebox{\myy}{
\rotatebox[origin=c]{218}{$
\begin{tikzpicture}[line width=1.5pt, color=gray!165]
    \draw [-{Classical TikZ Rightarrow[length=\myarrowlen]}](0,0) -- (0.05,0);
\end{tikzpicture}
$}}
\hspace{\myx}
\def \myx {226.8pt}
\def \myy {-4pt}
\hspace{-\myx}
\raisebox{\myy}{
\rotatebox[origin=c]{90}{$
\begin{tikzpicture}[line width=1.5pt, color=gray!165]
    \draw [-{Classical TikZ Rightarrow[length=\myarrowlen]}](0,0) -- (0.05,0);
\end{tikzpicture}
$}}
\hspace{\myx}
\def \myx {284.3pt}
\def \myy {-4pt}
\hspace{-\myx}
\raisebox{\myy}{
\rotatebox[origin=c]{90}{$
\begin{tikzpicture}[line width=1.5pt, color=gray!165]
    \draw [-{Classical TikZ Rightarrow[length=\myarrowlen]}](0,0) -- (0.05,0);
\end{tikzpicture}
$}}
\hspace{\myx}
\def \myx {251.5pt}
\def \myy {10pt}
\hspace{-\myx}
\raisebox{\myy}{
\rotatebox[origin=c]{90}{$
\begin{tikzpicture}[line width=1.5pt, color=gray!165]
    \draw [-{Classical TikZ Rightarrow[length=\myarrowlen]}](0,0) -- (0.05,0);
\end{tikzpicture}
$}}
\hspace{\myx}
\def \myx {299.5pt}
\def \myy {-18.8pt}
\hspace{-\myx}
\raisebox{\myy}{
\rotatebox[origin=c]{90}{$
\begin{tikzpicture}[line width=1.5pt, color=gray!165]
    \draw [-{Classical TikZ Rightarrow[length=\myarrowlen]}](0,0) -- (0.05,0);
\end{tikzpicture}
$}}
\hspace{\myx}
\hspace{\mycompensate}
\hspace{-245pt}
}

%% file: sections/F_doublesym_luca.tex
\section{Doubling the symmetry, the \ztwo case}
\label{sec:doublesym}
We are now in the position to  introduce a novel \rrr{symmetric} gauge-invariant tensor network formulation of lattice gauge theory. We start by providing the explicit example for the simplest case of the \ztwo lattice gauge theory, defined on the standard Kogut-Susskind group algebra. It is important to note that the copy tensors used in the standard PEPS LGT ansatz, i.e. Eq.~\eqref{eq:ansatz:Copyten}, do not fulfill the \ztwo sum-rule from Eq.~\eqref{eq:TN:Ccompo}. In this section, we demonstrate a method to modify the system, enabling the use of the same \textit{copying} strategy while producing globally symmetric tensors compatible with \rrr{standard} TN libraries' sum-rules.

We start by introducing spurious constituents on every link of the lattice. These constituents have Hilbert spaces that share the same charges as the original link constituents but with a degeneracy 1 for each irreducible representation. Consequently, an additional \ztwo charge is introduced at each site, leading to a symmetric basis $\{\ket{pq{{;}\alpha_{pq}}}\}$, where $p,q{\in}\{0{,}1\}$ denote the charges and $\alpha_{pq}$ the degeneracy label\footnote{If we identify the original charges with $p$ and the new spurious one with $q$ notice that given that the spurious constituents have degeneracy one in every irrep by construction $\alpha_{pq}\equiv \alpha_p$.}. Furthermore, the auxiliary legs of the TN are similarly composed of these \textit{doubled} vector spaces.

On each constituents on the links of the original lattice we map the original Hilbert space into the new doubled Hilbert space with the following identifications:
\begin{gather}
    \ket{0{;}1}\longleftrightarrow\ket{00{;}1}
    \qquad\text{and}\qquad\ket{1{;}1}\longleftrightarrow\ket{11{;}1}\ .
    \label{eq:double:phy_map}
\end{gather}
Since this extension is an embedding of our original Hilbert space into a larger one, we also need to embed the operators we use to define the \ham of the system.
Therefore, operators also change according to Eq.\eqref{eq:double:phy_map}, e.g., the $\sigma_x^l$ in this doubled system is mapped to $\ket{00{;}1}\bra{00{;}1}-\ket{11{;}1}\bra{11{;}1}$. The orthogonal complement to the charge-symmetric subspace remains unaffected by the dynamics and, as a result, completely decouples from the symmetric subspace. These components can therefore be safely discarded in the calculations. Notice that with our procedure we have enlarged the \rrr{considered} symmetry from the original \ztwo to a \zz symmetry on every link. The gauge-invariant states are then built as appropriate projectors from the symmetric subspace of such \ztwo to a \zz Hilbert space of constituents.

We are now in the position to describe our new gauge invariant tensor network  ansatz. We start by defining the following tensors, for all $i\in\mathcal{V}$ and $l\in\mathcal{L}$:
\begin{itemize}[leftmargin=*]
    \item[$*$] $p$-type vertex tensors $V_i$ ($p_\mathrm{tot}{=}g_i$ and $q_\mathrm{tot}{=}0$)
\end{itemize}
\begin{gather}
    \hspace{-6pt}
    \input{editions/pvertex_ten}
    =
    \begin{cases}
        \left[V_i\right]^{a0{,}b0{,}c0{,}d0}_{\alpha\beta\gamma\delta} &
        \parbox[t]{.135\textwidth}{\raggedleft
        if $a{+}b{+}c{+}d{=}g_i$ $\modtwo$
        }
        \\
        0 & \text{otherwise}
    \end{cases}
    \def \myshift {60pt}
    \def \mybracelen {56pt}
    \hspace{-\myshift}
    \raisebox{19pt}{$\overbrace{\hspace{\mybracelen}}^\text{Gauss' law on $p$}$}
    \ ,
    \label{eq:double:pvertex}
\end{gather} 
\begin{itemize}[leftmargin=*]
    \item[$*$] $q$-type vertex tensors $V_i$ ($p_\mathrm{tot}{=}0$ and $q_\mathrm{tot}{=}g_i$)
\end{itemize}
\begin{gather}
    \hspace{-6pt}
    \input{editions/qvertex_ten}
    =
    \begin{cases}
        \left[V_i\right]^{0a{,}0b{,}0c{,}0d}_{\alpha\beta\gamma\delta} &
        \parbox[t]{.135\textwidth}{\raggedleft
        if $a{+}b{+}c{+}d{=}g_i$ $\modtwo$
        }
        \\
        0 & \text{otherwise}
    \end{cases}
    \def \myshift {60pt}
    \def \mybracelen {56pt}
    \hspace{-\myshift}
    \raisebox{19pt}{$\overbrace{\hspace{\mybracelen}}^\text{Gauss' law on $q$}$}
    \ ,
    \label{eq:double:qvertex}
\end{gather}
\begin{itemize}[leftmargin=*]
    \item[$*$] $pq$-type copy tensors $C_l$ ($p_\mathrm{tot}{=}0$ and $q_\mathrm{tot}{=}0$)
\end{itemize}
\begin{gather}
    \input{editions/pqcopy_ten}
    =\begin{cases}
        \left[C_l\right]^{a0{,}bb{,}0c}_{\alpha 1\gamma} & \text{if }a=b=c \\
        0 & \text{otherwise}
    \end{cases}
    \def \myshift {43pt}
    \def \mybracelen {40pt}
    \hspace{-\myshift}
    \raisebox{15pt}{$\overbrace{\hspace{\mybracelen}}^\text{copy charge}$}
    \ ,
    \label{eq:double:pqcopy}
\end{gather}
\begin{itemize}[leftmargin=*]
    \item[$*$] $qp$-type copy tensors $C_l$ ($p_\mathrm{tot}{=}0$ and $q_\mathrm{tot}{=}0$)
\end{itemize}
\begin{gather}
    \input{editions/qpcopy_ten}
    =\begin{cases}
        \left[C_l\right]^{0a{,}bb{,}c0}_{\alpha 1\gamma} & \text{if }a=b=c \\
        0 & \text{otherwise}
    \end{cases}
    \def \myshift {43pt}
    \def \mybracelen {40pt}
    \hspace{-\myshift}
    \raisebox{15pt}{$\overbrace{\hspace{\mybracelen}}^\text{copy charge}$}
    \ .
    \label{eq:double:qpcopy}
\end{gather}

The labels $p, q, pq,qp$ of the tensors refer to which charge among the two $G\times G$ charges  \textit{carries} the physical information. For instance, in Eq.~\eqref{eq:double:qpcopy}, the copy tensor copies the physical charge onto the second charge ($q$) of the left leg and the first charge ($p$) of the right leg. Hence, $qp$-type copy tensor. Another example is found in Eq.~\eqref{eq:double:qvertex}, where all physical information resides on the second charge ($q$), i.e. $q$-type vertex tensor. It is precisely through this charge that Gauss' law is imposed. Finally, the layout of the modified ansatz is
\begin{equation}
    \ket\Psi=
    \input{editions/layoutz2z2.tex}\ .
    \label{eq:gral:layout2nz2}
\end{equation}

Since we are successfully using the same \textit{copying} strategy, one can check, similar to Eq.~\eqref{eq:ansatz:checkvertex}, that this modified ansatz still exclusively represents locally symmetric states. Crucially, the tensors are now \zz globally symmetric, i.e. they fulfill the \ztwo sum-rule on \textit{both} charges \textit{simultaneously}. For example, consider the components of a $pq$-type copy tensor from Eq.~\eqref{eq:double:pqcopy} and the sum-rules on each charge:
\begin{align}
    [C_l]^{00{,}00{,}00}_{\alpha 1\gamma}
    \xrightarrow{}
    \begin{Bmatrix}
        p:\ 0{+}0{+}0\modtwo{=}0\\
        q:\ 0{+}0{+}0\modtwo{=}0
    \end{Bmatrix}
    \implies
    \begin{matrix}
        p_\mathrm{tot}{=}0\\
        q_\mathrm{tot}{=}0
    \end{matrix}
    \ \texttransparent{0}{.}
    \nonumber\\
    [C_l]^{10{,}11{,}01}_{\alpha 1\gamma}
    \xrightarrow{}
    \begin{Bmatrix}
        p:\ 1{+}1{+}0\modtwo{=}0\\
        q:\ 0{+}1{+}1\modtwo{=}0
    \end{Bmatrix}
    \implies
    \begin{matrix}
        p_\mathrm{tot}{=}0\\
        q_\mathrm{tot}{=}0
    \end{matrix}
    \ .
\end{align}
Both sectors are given by sum-rules with $p_\mathrm{tot}{=}q_\mathrm{tot}{=}0$, meaning $C_l$ is \zz symmetric! Similarly for the rest of the TN: tensors are \zz symmetric because components are compatible with sum-rules using the given total charges.

In summary, we mapped the \ztwo locally symmetric ansatz to a subspace of the \zz globally symmetric states on a doubled system. Furthermore, since the tensors are globally \zz symmetric, they can be efficiently handled by globally symmetric TN libraries. In other words, we can now profit from the block diagonal structure of the elementary tensors and thus  efficiently simulate locally symmetric states with standard  TN libraries.

In the next section we will elaborate more on the generic ansatz for \zn gauge theories.

%% file: editions/pvertex_ten.tex
\ensuremath{
\begin{matrix}
    \texttransparent{1}{
    \includegraphics[scale=0.108]{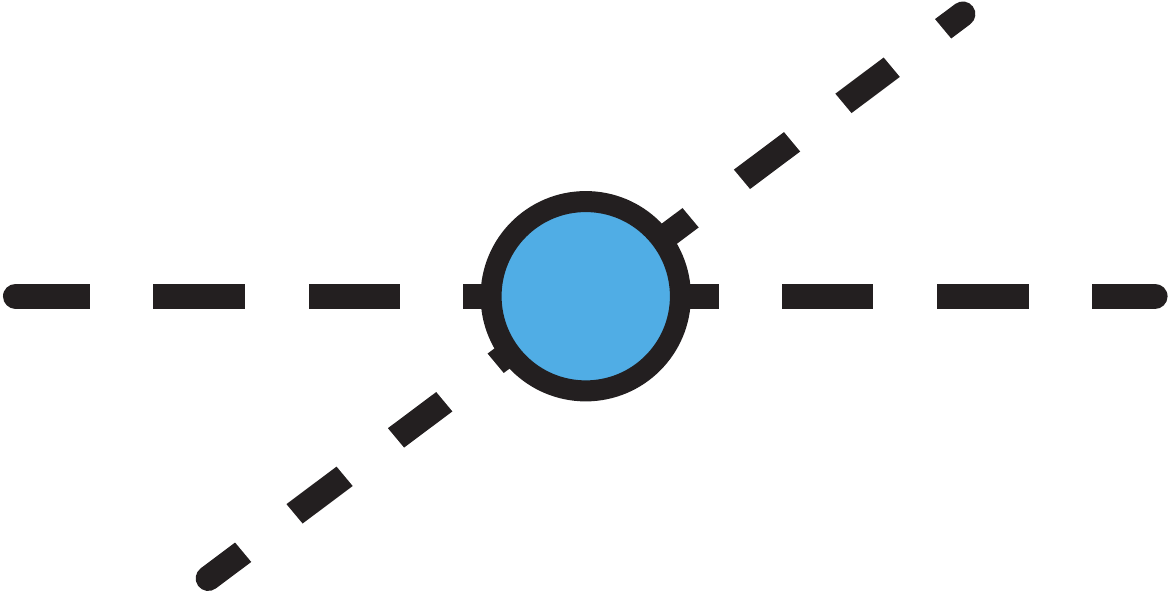}
    }
\end{matrix}
\def \myx {25pt}
\def \myy {25pt}
\hspace{-\myx}
\raisebox{\myy}{
$\ket{a0{;}\alpha}$
}
\hspace{\myx}
\def \myx {100pt}
\def \myy {-22pt}
\hspace{-\myx}
\raisebox{\myy}{
$\ket{c0{;}\gamma}$
}
\hspace{\myx}
\def \myx {90pt}
\def \myy {-7pt}
\hspace{-\myx}
\raisebox{\myy}{
$\ket{b0{;}\beta}$
}
\hspace{\myx}
\def \myx {165pt}
\def \myy {10pt}
\hspace{-\myx}
\raisebox{\myy}{
$\ket{d0{;}\delta}$
}
\hspace{\myx}
\def \myx {165pt}
\def \myy {15pt}
\hspace{-\myx}
\raisebox{\myy}{
$V_i$
}
\hspace{\myx}
\def \myx {180pt}
\def \myy {-10pt}
\hspace{-\myx}
\raisebox{\myy}{
$g_i$
}
\hspace{\myx}
\def \myx {169pt}
\def \myy {1.0pt}
\hspace{-\myx}
\raisebox{\myy}{$
\begin{tikzpicture}[line width=1.5pt, color=gray!165]
    \draw [-{Classical TikZ Rightarrow[length=3.5pt]}](0,0) -- (0.05,0);
\end{tikzpicture}
$}
\hspace{\myx}
\def \myx {214pt}
\def \myy {1.0pt}
\hspace{-\myx}
\raisebox{\myy}{$
\begin{tikzpicture}[line width=1.5pt, color=gray!165]
    \draw [-{Classical TikZ Rightarrow[length=3.5pt]}](0,0) -- (0.05,0);
\end{tikzpicture}
$}
\hspace{\myx}
\def \myx {189.5pt}
\def \myy {11.5pt}
\hspace{-\myx}
\raisebox{\myy}{
\rotatebox[origin=c]{218}{$
\begin{tikzpicture}[line width=1.5pt, color=gray!165]
    \draw [-{Classical TikZ Rightarrow[length=3.5pt]}](0,0) -- (0.05,0);
\end{tikzpicture}
$}}
\hspace{\myx}
\def \myx {228.5pt}
\def \myy {-9.5pt}
\hspace{-\myx}
\raisebox{\myy}{
\rotatebox[origin=c]{218}{$
\begin{tikzpicture}[line width=1.5pt, color=gray!165]
    \draw [-{Classical TikZ Rightarrow[length=3.5pt]}](0,0) -- (0.05,0);
\end{tikzpicture}
$}}
\hspace{\myx}
\hspace{-185pt}
}

%% file: editions/qvertex_ten.tex
\ensuremath{
\begin{matrix}
    \includegraphics[scale=1.1]{images/4leg.pdf}
\end{matrix}
\texttransparent{1}{\llap{$
\begin{matrix}
    \includegraphics[trim= 37.8 0 0 0, scale=1.11]{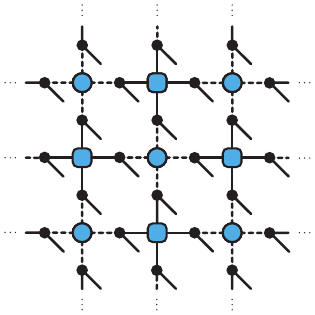}
\end{matrix}
$}}
\def \myx {25pt}
\def \myy {25pt}
\hspace{-\myx}
\raisebox{\myy}{
$\ket{0a{;}\alpha}$
}
\hspace{\myx}
\def \myx {100pt}
\def \myy {-22pt}
\hspace{-\myx}
\raisebox{\myy}{
$\ket{0c{;}\gamma}$
}
\hspace{\myx}
\def \myx {90pt}
\def \myy {-7pt}
\hspace{-\myx}
\raisebox{\myy}{
$\ket{0b{;}\beta}$
}
\hspace{\myx}
\def \myx {165pt}
\def \myy {10pt}
\hspace{-\myx}
\raisebox{\myy}{
$\ket{0d{;}\delta}$
}
\hspace{\myx}
\def \myx {165pt}
\def \myy {15pt}
\hspace{-\myx}
\raisebox{\myy}{
$V_i$
}
\hspace{\myx}
\def \myx {180pt}
\def \myy {-10pt}
\hspace{-\myx}
\raisebox{\myy}{
$g_i$
}
\hspace{\myx}
\def \mycompensate {0pt}
\hspace{-\mycompensate}
\def \myx {169pt}
\def \myy {1.0pt}
\hspace{-\myx}
\raisebox{\myy}{$
\begin{tikzpicture}[line width=1.5pt, color=gray!165]
    \draw [-{Classical TikZ Rightarrow[length=3.5pt]}](0,0) -- (0.05,0);
\end{tikzpicture}
$}
\hspace{\myx}
\def \myx {214pt}
\def \myy {1.0pt}
\hspace{-\myx}
\raisebox{\myy}{$
\begin{tikzpicture}[line width=1.5pt, color=gray!165]
    \draw [-{Classical TikZ Rightarrow[length=3.5pt]}](0,0) -- (0.05,0);
\end{tikzpicture}
$}
\hspace{\myx}
\def \myx {189.5pt}
\def \myy {11.5pt}
\hspace{-\myx}
\raisebox{\myy}{
\rotatebox[origin=c]{218}{$
\begin{tikzpicture}[line width=1.5pt, color=gray!165]
    \draw [-{Classical TikZ Rightarrow[length=3.5pt]}](0,0) -- (0.05,0);
\end{tikzpicture}
$}}
\hspace{\myx}
\def \myx {228.5pt}
\def \myy {-9.5pt}
\hspace{-\myx}
\raisebox{\myy}{
\rotatebox[origin=c]{218}{$
\begin{tikzpicture}[line width=1.5pt, color=gray!165]
    \draw [-{Classical TikZ Rightarrow[length=3.5pt]}](0,0) -- (0.05,0);
\end{tikzpicture}
$}}
\hspace{\myx}
\hspace{\mycompensate}
\hspace{-185pt}
}

%% file: editions/pqcopy_ten.tex
\ensuremath{
\begin{matrix}
    \texttransparent{0.93}{
    \includegraphics[scale=0.0175]{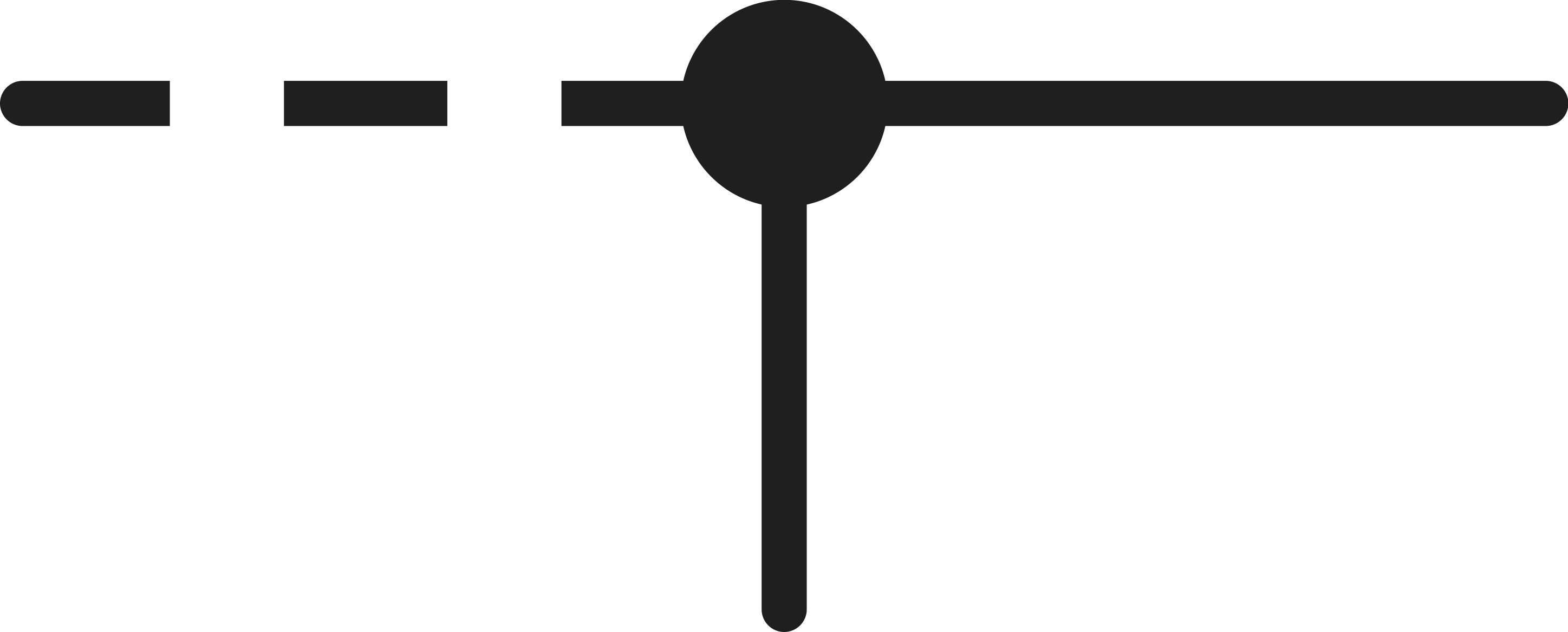}
    }
\end{matrix}
\def \myx {66pt}
\def \myy {0pt}
\hspace{-\myx}
\raisebox{\myy}{
$\ket{a0{;}\alpha}$
}
\hspace{\myx}
\def \myx {47pt}
\def \myy {0pt}
\hspace{-\myx}
\raisebox{\myy}{
$\ket{0c{;}\gamma}$
}
\hspace{\myx}
\def \myx {105pt}
\def \myy {-16pt}
\hspace{-\myx}
\raisebox{\myy}{
$\ket{bb{;}1}$
}
\hspace{\myx}
\def \myx {127pt}
\def \myy {18pt}
\hspace{-\myx}
\raisebox{\myy}{
$C_l$
}
\hspace{\myx}
\hspace{0.7pt}
\def \myx {125pt}
\def \myy {7.8pt}
\hspace{-\myx}
\raisebox{\myy}{$
\begin{tikzpicture}[line width=1.5pt, color=gray!165]
    \draw [-{Classical TikZ Rightarrow[length=3.5pt]}](0,0) -- (0.05,0);
\end{tikzpicture}
$}
\hspace{\myx}
\def \myx {156pt}
\def \myy {7.8pt}
\hspace{-\myx}
\raisebox{\myy}{$
\begin{tikzpicture}[line width=1.5pt, color=gray!165]
    \draw [-{Classical TikZ Rightarrow[length=3.5pt]}](0,0) -- (0.05,0);
\end{tikzpicture}
$}
\hspace{\myx}
\def \myx {152.1pt}
\def \myy {-2pt}
\hspace{-\myx}
\raisebox{\myy}{
\rotatebox[origin=c]{90}{$
\begin{tikzpicture}[line width=1.5pt, color=gray!165]
    \draw [-{Classical TikZ Rightarrow[length=3.5pt]}](0,0) -- (0.05,0);
\end{tikzpicture}
$}}
\hspace{\myx}
\hspace{-115pt}
}

%% file: editions/qpcopy_ten.tex
\ensuremath{
\begin{matrix}
    \texttransparent{0.93}{
    \includegraphics[scale=0.0175]{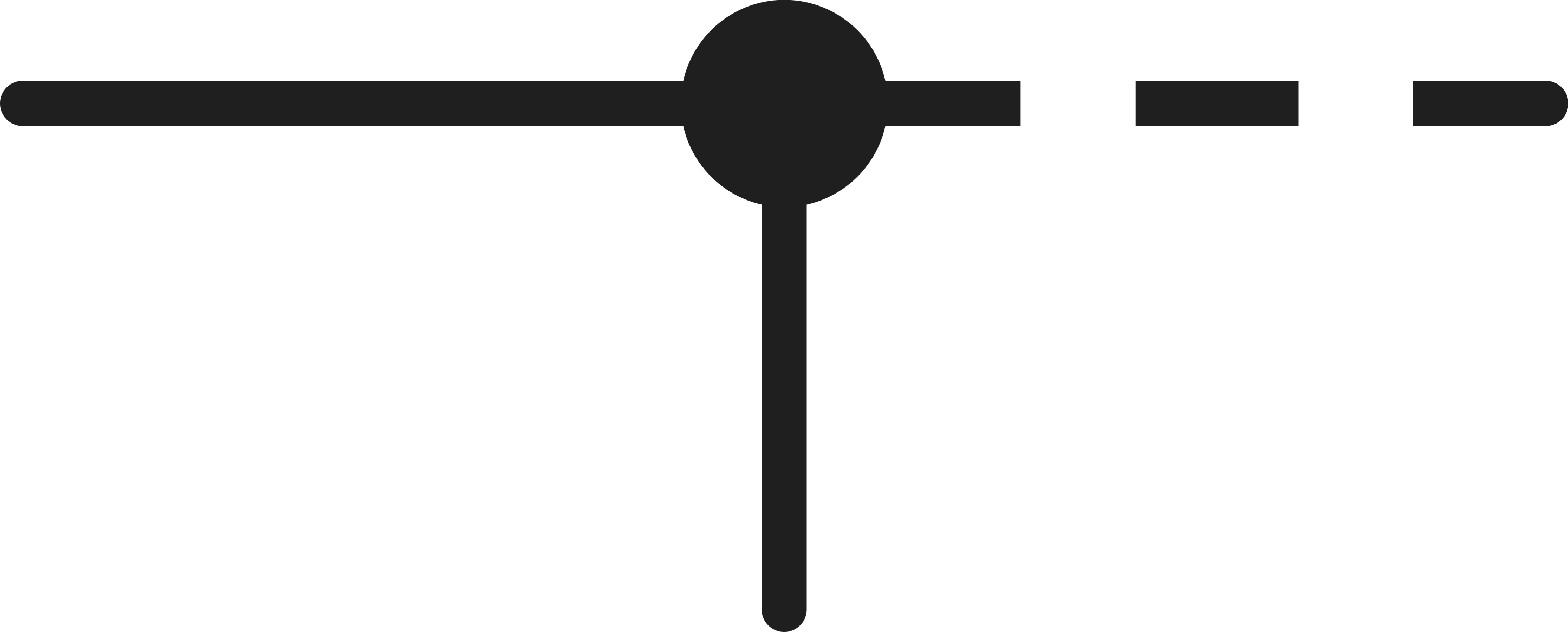}
    }
\end{matrix}
\def \myx {66pt}
\def \myy {0pt}
\hspace{-\myx}
\raisebox{\myy}{
$\ket{0a{;}\alpha}$
}
\hspace{\myx}
\def \myx {47pt}
\def \myy {0pt}
\hspace{-\myx}
\raisebox{\myy}{
$\ket{c0{;}\gamma}$
}
\hspace{\myx}
\def \myx {105pt}
\def \myy {-16pt}
\hspace{-\myx}
\raisebox{\myy}{
$\ket{bb{;}1}$
}
\hspace{\myx}
\def \myx {127pt}
\def \myy {18pt}
\hspace{-\myx}
\raisebox{\myy}{
$C_l$
}
\hspace{\myx}
\hspace{0.7pt}
\def \myx {125pt}
\def \myy {7.5pt}
\hspace{-\myx}
\raisebox{\myy}{$
\begin{tikzpicture}[line width=1.5pt, color=gray!165]
    \draw [-{Classical TikZ Rightarrow[length=3.5pt]}](0,0) -- (0.05,0);
\end{tikzpicture}
$}
\hspace{\myx}
\def \myx {156pt}
\def \myy {7.5pt}
\hspace{-\myx}
\raisebox{\myy}{$
\begin{tikzpicture}[line width=1.5pt, color=gray!165]
    \draw [-{Classical TikZ Rightarrow[length=3.5pt]}](0,0) -- (0.05,0);
\end{tikzpicture}
$}
\hspace{\myx}
\def \myx {152.1pt}
\def \myy {-2pt}
\hspace{-\myx}
\raisebox{\myy}{
\rotatebox[origin=c]{90}{$
\begin{tikzpicture}[line width=1.5pt, color=gray!165]
    \draw [-{Classical TikZ Rightarrow[length=3.5pt]}](0,0) -- (0.05,0);
\end{tikzpicture}
$}}
\hspace{\myx}
\hspace{-115pt}
}

%% file: editions/layoutz2z2.tex
\ensuremath{
\mypicnorm[0pt][1.1]{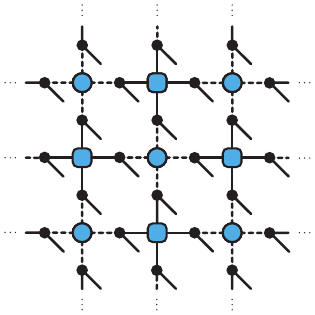}
\def \hcorr {7.7pt}
\def \vcorr {10.0815pt}
\def \dcorr {11.01pt}
\def \firstcolx {129.25pt}
\def \secondcolx {89.3pt}
\def \thirdcolx {49.6pt}
\def \firstrowy {40.45pt}
\def \secondrowy {0.81pt}
\def \thirdrowy {-39.08pt}
\def \siteangle {135}
\myarrow{154.5pt}{\firstrowy}{0}{-\hcorr}
\myarrow{138.5pt}{\firstrowy}{0}{-\hcorr}
\myarrow{118.6pt}{\firstrowy}{0}{-\hcorr}
\myarrow{99.5pt}{\firstrowy}{0}{-\hcorr}
\myarrow{76.7pt}{\firstrowy}{0}{-\hcorr}
\myarrow{58.5pt}{\firstrowy}{0}{-\hcorr}
\myarrow{38.5pt}{\firstrowy}{0}{-\hcorr}
\myarrow{21.5pt}{\firstrowy}{0}{-\hcorr}
\myarrow{156.5pt}{\secondrowy}{0}{-\hcorr}
\myarrow{138.5pt}{\secondrowy}{0}{-\hcorr}
\myarrow{115.6pt}{\secondrowy}{0}{-\hcorr}
\myarrow{97.5pt}{\secondrowy}{0}{-\hcorr}
\myarrow{77.7pt}{\secondrowy}{0}{-\hcorr}
\myarrow{58.5pt}{\secondrowy}{0}{-\hcorr}
\myarrow{35.5pt}{\secondrowy}{0}{-\hcorr}
\myarrow{18.5pt}{\secondrowy}{0}{-\hcorr}
\hspace{2pt}
\myarrow{154.5pt}{\thirdrowy}{0}{-\hcorr}
\myarrow{138.5pt}{\thirdrowy}{0}{-\hcorr}
\myarrow{118.6pt}{\thirdrowy}{0}{-\hcorr}
\myarrow{99.5pt}{\thirdrowy}{0}{-\hcorr}
\myarrow{76.7pt}{\thirdrowy}{0}{-\hcorr}
\myarrow{58.5pt}{\thirdrowy}{0}{-\hcorr}
\myarrow{38.5pt}{\thirdrowy}{0}{-\hcorr}
\myarrow{21.5pt}{\thirdrowy}{0}{-\hcorr}
\myarrow{\firstcolx}{67pt}{-90}{-\vcorr}
\myarrow{\firstcolx}{50.5pt}{-90}{-\vcorr}
\myarrow{\firstcolx}{30.5pt}{-90}{-\vcorr}
\myarrow{\firstcolx}{12pt}{-90}{-\vcorr}
\myarrow{\firstcolx}{-10.5pt}{-90}{-\vcorr}
\myarrow{\firstcolx}{-29pt}{-90}{-\vcorr}
\myarrow{\firstcolx}{-49pt}{-90}{-\vcorr}
\myarrow{\firstcolx}{-65pt}{-90}{-\vcorr}
\myarrow{\secondcolx}{67pt}{-90}{-\vcorr}
\myarrow{\secondcolx}{50.5pt}{-90}{-\vcorr}
\myarrow{\secondcolx}{30.5pt}{-90}{-\vcorr}
\myarrow{\secondcolx}{12pt}{-90}{-\vcorr}
\myarrow{\secondcolx}{-10.5pt}{-90}{-\vcorr}
\myarrow{\secondcolx}{-29pt}{-90}{-\vcorr}
\myarrow{\secondcolx}{-49pt}{-90}{-\vcorr}
\myarrow{\secondcolx}{-68pt}{-90}{-\vcorr}
\myarrow{\thirdcolx}{67pt}{-90}{-\vcorr}
\myarrow{\thirdcolx}{50.5pt}{-90}{-\vcorr}
\myarrow{\thirdcolx}{30.5pt}{-90}{-\vcorr}
\myarrow{\thirdcolx}{12pt}{-90}{-\vcorr}
\myarrow{\thirdcolx}{-10.5pt}{-90}{-\vcorr}
\myarrow{\thirdcolx}{-29pt}{-90}{-\vcorr}
\myarrow{\thirdcolx}{-49pt}{-90}{-\vcorr}
\myarrow{\thirdcolx}{-65pt}{-90}{-\vcorr}
\myarrow{123pt}{54pt}{\siteangle}{-\dcorr}
\myarrow{123pt}{14pt}{\siteangle}{-\dcorr}
\myarrow{123pt}{-25.8pt}{\siteangle}{-\dcorr}
\myarrow{123pt}{-65.45pt}{\siteangle}{-\dcorr}
\myarrow{83.7pt}{54pt}{\siteangle}{-\dcorr}
\myarrow{83.7pt}{14.2pt}{\siteangle}{-\dcorr}
\myarrow{83.7pt}{-25.7pt}{\siteangle}{-\dcorr}
\myarrow{83.7pt}{-65.45pt}{\siteangle}{-\dcorr}
\myarrow{44.4pt}{54pt}{\siteangle}{-\dcorr}
\myarrow{44.4pt}{14.2pt}{\siteangle}{-\dcorr}
\myarrow{44.4pt}{-25.7pt}{\siteangle}{-\dcorr}
\myarrow{44.4pt}{-65.45pt}{\siteangle}{-\dcorr}
\myarrow{144pt}{34pt}{\siteangle}{-\dcorr}
\myarrow{104.4pt}{34pt}{\siteangle}{-\dcorr}
\myarrow{65pt}{34pt}{\siteangle}{-\dcorr}
\myarrow{24.8pt}{34pt}{\siteangle}{-\dcorr}
\myarrow{144pt}{-6pt}{\siteangle}{-\dcorr}
\myarrow{104.4pt}{-6pt}{\siteangle}{-\dcorr}
\myarrow{65pt}{-6pt}{\siteangle}{-\dcorr}
\myarrow{24.8pt}{-6pt}{\siteangle}{-\dcorr}
\myarrow{144pt}{-46pt}{\siteangle}{-\dcorr}
\myarrow{104.4pt}{-46pt}{\siteangle}{-\dcorr}
\myarrow{65pt}{-46pt}{\siteangle}{-\dcorr}
\myarrow{24.8pt}{-46pt}{\siteangle}{-\dcorr}
}

%% file: sections/GA_generalisation_luca.tex
\section{Tensor network ansatz for \znzn using elementary symmetric tensors}
\label{sec:general}

With the basic \zn concepts presented in Sec.~\ref{sec:gauge} in mind, we now present the generalisation of the new tensor network  ansatz for  \zn gauge theories. As with the \ztwo case, we begin by doubling each original vector space on the ansatz from $\mathbb{W}$ to $\mathbb{W}\otimes\mathbb{W}$. This leads to a new diagonal basis, $\{\ket{pq;\alpha_{pq}}\}$, analogous to the \ztwo construction. \rrr{However, in this generalized approach, a new state identification is established between the original Hilbert space and its doubled counterpart.} We identify the states on physical sites\footnote{Label the vertex $i$ of a square lattice as $i{=}(x{,}y)$ with integers $x, y$. The parity of the \textit{right} and \textit{bottom} sites of the vertex is given by the parity of $x{+}y$.} as follows:
\begin{align}
    \text{\textit{even} sites:}&\quad
    \ket{p; 1}
    \quad\longleftrightarrow\quad
    \ket{p(-p); 1}, \nonumber\\
    \text{\textit{odd} sites:}&\quad
    \ket{p; 1}
    \quad\longleftrightarrow\quad
    \ket{(-p)p; 1},
    \label{eq:gral:identify}
\end{align}
with $p\in\{0,\dots ,N-1\}$. It follows from the definition of $\ket{p}$ that $\ket{(-p);1}{\equiv}\ket{N{-}p;1}$, since both relate to the unique eigenvector with eigenvalue \rrr{$\exp\left(i\frac{2\pi}{N}(-p)\right)=\exp\left(i\frac{2\pi}{N}(N-p)\right)$}. Operators acting on physical sites change accordingly. As an example, in the $\mathbb{Z}_3$ case on an even site, an operator would undergo to the following transformation (omitting degeneracy):
\begin{gather}
    c_0\ket{0}\bra{0}+
    c_1\ket{1}\bra{1}+
    c_2\ket{2}\bra{2}\nonumber\\
    \updownarrow\\
    c_0\ket{00}\bra{00}+
    c_1\ket{12}\bra{12}+
    c_2\ket{21}\bra{21}.\nonumber
    \label{eq:gral:exZ3}
\end{gather}

The \znzn-symmetric ansatz on these doubled vector spaces has the same layout as the \zz case, which we give again
\begin{equation}
    \ket\Psi=
    \input{editions/layoutznzn.tex}\ .
    \label{eq:gral:layoutznzn}
\end{equation}
Its new elementary tensors are the following:
\begin{itemize}[leftmargin=*]
    \item[$*$] $p$-type vertex tensors $V_i$ ($p_\mathrm{tot}{=}g_i$ and $q_\mathrm{tot}{=}\rrr{0}$)
\end{itemize}
\begin{gather}
    \hspace{-6pt}
    \input{editions/pvertex_ten_znzn}
    =
    \begin{cases}
        \left[V_i\right]^{a0{,}b0{,}c0{,}d0}_{\alpha\beta\gamma\delta} &
        \parbox[t]{.135\textwidth}{\raggedleft
        if $a{+}d{-}b{-}c{=}g_i$ $\modn$
        }
        \\
        0 & \text{otherwise}
    \end{cases}
    \def \myshift {60pt}
    \def \mybracelen {56pt}
    \hspace{-\myshift}
    \raisebox{19pt}{$\overbrace{\hspace{\mybracelen}}^\text{Gauss' law on $p$}$}
    \ ,
    \label{eq:gral:pvertex}
\end{gather} 
\begin{itemize}[leftmargin=*]
    \item[$*$] $q$-type vertex tensors $V_i$ ($p_\mathrm{tot}{=}\rrr{0}$ and $q_\mathrm{tot}{=}g_i$)
\end{itemize}
\begin{gather}
    \hspace{-6pt}
    \input{editions/qvertex_ten_znzn}
    =
    \begin{cases}
        \left[V_i\right]^{0a{,}0b{,}0c{,}0d}_{\alpha\beta\gamma\delta} &
        \parbox[t]{.135\textwidth}{\raggedleft
        if $a{+}d{-}b{-}c{=}g_i$ $\modn$
        }
        \\
        0 & \text{otherwise}
    \end{cases}
    \def \myshift {60pt}
    \def \mybracelen {56pt}
    \hspace{-\myshift}
    \raisebox{19pt}{$\overbrace{\hspace{\mybracelen}}^\text{Gauss' law on $q$}$}
    \ ,
    \label{eq:gral:qvertex}
\end{gather}
\begin{itemize}[leftmargin=*]
    \item[$*$] $pq$-type copy tensors $C_l$ ($p_\mathrm{tot}{=}\rrr{0}$ and $q_\mathrm{tot}{=}\rrr{0}$)
\end{itemize}
\begin{gather}
    \input{editions/pqcopy_ten_znzn}
    =\begin{cases}
        \left[C_l\right]^{a0{,}(-b)b{,}0c}_{\alpha 1\gamma} & \text{if }a=b=c \\
        0 & \text{otherwise}
    \end{cases}
    \def \myshift {43pt}
    \def \mybracelen {40pt}
    \hspace{-\myshift}
    \raisebox{15pt}{$\overbrace{\hspace{\mybracelen}}^\text{copy charge}$}
    \ ,
    \label{eq:gral:pqcopy}
\end{gather}
\begin{itemize}[leftmargin=*]
    \item[$*$] $qp$-type copy tensors $C_l$ ($p_\mathrm{tot}{=}\rrr{0}$ and $q_\mathrm{tot}{=}\rrr{0}$)
\end{itemize}
\begin{gather}
    \input{editions/qpcopy_ten_znzn}
    =\begin{cases}
        \left[C_l\right]^{0a{,}b(-b){,}c0}_{\alpha 1\gamma} & \text{if }a=b=c \\
        0 & \text{otherwise}
    \end{cases}
    \def \myshift {43pt}
    \def \mybracelen {40pt}
    \hspace{-\myshift}
    \raisebox{15pt}{$\overbrace{\hspace{\mybracelen}}^\text{copy charge}$}
    \ .
    \label{eq:gral:qpcopy}
\end{gather}

Similar to the \ztwo case, it can be verified that the states $\ket\Psi$ represented by the ansatz from Eq.~\eqref{eq:gral:layoutznzn} satisfy the \zn-Gauss' law. Furthermore, all tensors are \znzn symmetric, as they fulfill Eq.~\eqref{eq:inv_tensor} for \textit{each} charge. Note that, as shown in the layout of Eq.~\eqref{eq:gral:layoutznzn}, all $pq$-type copy tensors are positioned on odd sites, while all $qp$-type copy tensors are placed on even sites (or vice-versa), aligning with the identification in Eq.~\eqref{eq:gral:identify}. 

In conclusion, the TN described by \crefrange{eq:gral:layoutznzn}{eq:gral:qpcopy} generalises the ansatz to \znzn. Essentially, a new identification of states in physical sites has been introduced, which reproduces the \ztwo case when $N{=}2$. \rrr{Beyond this, the \zz ansatz extends naturally to larger $N$.}

%% file: editions/layoutznzn.tex
\ensuremath{
\mypicnorm[0pt][1.1]{images/checkers.pdf}
\def \hcorr {7.7pt}
\def \vcorr {10.0815pt}
\def \dcorr {11.01pt}
\def \firstcolx {129.25pt}
\def \secondcolx {89.3pt}
\def \thirdcolx {49.6pt}
\def \firstrowy {40.45pt}
\def \secondrowy {0.81pt}
\def \thirdrowy {-39.08pt}
\def \siteangle {135}
\myarrow{154.5pt}{\firstrowy}{0}{-\hcorr}
\myarrow{138.5pt}{\firstrowy}{0}{-\hcorr}
\myarrow{118.6pt}{\firstrowy}{0}{-\hcorr}
\myarrow{99.5pt}{\firstrowy}{0}{-\hcorr}
\myarrow{76.7pt}{\firstrowy}{0}{-\hcorr}
\myarrow{58.5pt}{\firstrowy}{0}{-\hcorr}
\myarrow{38.5pt}{\firstrowy}{0}{-\hcorr}
\myarrow{21.5pt}{\firstrowy}{0}{-\hcorr}
\myarrow{156.5pt}{\secondrowy}{0}{-\hcorr}
\myarrow{138.5pt}{\secondrowy}{0}{-\hcorr}
\myarrow{115.6pt}{\secondrowy}{0}{-\hcorr}
\myarrow{97.5pt}{\secondrowy}{0}{-\hcorr}
\myarrow{77.7pt}{\secondrowy}{0}{-\hcorr}
\myarrow{58.5pt}{\secondrowy}{0}{-\hcorr}
\myarrow{35.5pt}{\secondrowy}{0}{-\hcorr}
\myarrow{18.5pt}{\secondrowy}{0}{-\hcorr}
\hspace{2pt}
\myarrow{154.5pt}{\thirdrowy}{0}{-\hcorr}
\myarrow{138.5pt}{\thirdrowy}{0}{-\hcorr}
\myarrow{118.6pt}{\thirdrowy}{0}{-\hcorr}
\myarrow{99.5pt}{\thirdrowy}{0}{-\hcorr}
\myarrow{76.7pt}{\thirdrowy}{0}{-\hcorr}
\myarrow{58.5pt}{\thirdrowy}{0}{-\hcorr}
\myarrow{38.5pt}{\thirdrowy}{0}{-\hcorr}
\myarrow{21.5pt}{\thirdrowy}{0}{-\hcorr}
\myarrow{\firstcolx}{67pt}{-90}{-\vcorr}
\myarrow{\firstcolx}{50.5pt}{-90}{-\vcorr}
\myarrow{\firstcolx}{30.5pt}{-90}{-\vcorr}
\myarrow{\firstcolx}{12pt}{-90}{-\vcorr}
\myarrow{\firstcolx}{-10.5pt}{-90}{-\vcorr}
\myarrow{\firstcolx}{-29pt}{-90}{-\vcorr}
\myarrow{\firstcolx}{-49pt}{-90}{-\vcorr}
\myarrow{\firstcolx}{-65pt}{-90}{-\vcorr}
\myarrow{\secondcolx}{67pt}{-90}{-\vcorr}
\myarrow{\secondcolx}{50.5pt}{-90}{-\vcorr}
\myarrow{\secondcolx}{30.5pt}{-90}{-\vcorr}
\myarrow{\secondcolx}{12pt}{-90}{-\vcorr}
\myarrow{\secondcolx}{-10.5pt}{-90}{-\vcorr}
\myarrow{\secondcolx}{-29pt}{-90}{-\vcorr}
\myarrow{\secondcolx}{-49pt}{-90}{-\vcorr}
\myarrow{\secondcolx}{-68pt}{-90}{-\vcorr}
\myarrow{\thirdcolx}{67pt}{-90}{-\vcorr}
\myarrow{\thirdcolx}{50.5pt}{-90}{-\vcorr}
\myarrow{\thirdcolx}{30.5pt}{-90}{-\vcorr}
\myarrow{\thirdcolx}{12pt}{-90}{-\vcorr}
\myarrow{\thirdcolx}{-10.5pt}{-90}{-\vcorr}
\myarrow{\thirdcolx}{-29pt}{-90}{-\vcorr}
\myarrow{\thirdcolx}{-49pt}{-90}{-\vcorr}
\myarrow{\thirdcolx}{-65pt}{-90}{-\vcorr}
\myarrow{123pt}{54pt}{\siteangle}{-\dcorr}
\myarrow{123pt}{14pt}{\siteangle}{-\dcorr}
\myarrow{123pt}{-25.8pt}{\siteangle}{-\dcorr}
\myarrow{123pt}{-65.45pt}{\siteangle}{-\dcorr}
\myarrow{83.7pt}{54pt}{\siteangle}{-\dcorr}
\myarrow{83.7pt}{14.2pt}{\siteangle}{-\dcorr}
\myarrow{83.7pt}{-25.7pt}{\siteangle}{-\dcorr}
\myarrow{83.7pt}{-65.45pt}{\siteangle}{-\dcorr}
\myarrow{44.4pt}{54pt}{\siteangle}{-\dcorr}
\myarrow{44.4pt}{14.2pt}{\siteangle}{-\dcorr}
\myarrow{44.4pt}{-25.7pt}{\siteangle}{-\dcorr}
\myarrow{44.4pt}{-65.45pt}{\siteangle}{-\dcorr}
\myarrow{144pt}{34pt}{\siteangle}{-\dcorr}
\myarrow{104.4pt}{34pt}{\siteangle}{-\dcorr}
\myarrow{65pt}{34pt}{\siteangle}{-\dcorr}
\myarrow{24.8pt}{34pt}{\siteangle}{-\dcorr}
\myarrow{144pt}{-6pt}{\siteangle}{-\dcorr}
\myarrow{104.4pt}{-6pt}{\siteangle}{-\dcorr}
\myarrow{65pt}{-6pt}{\siteangle}{-\dcorr}
\myarrow{24.8pt}{-6pt}{\siteangle}{-\dcorr}
\myarrow{144pt}{-46pt}{\siteangle}{-\dcorr}
\myarrow{104.4pt}{-46pt}{\siteangle}{-\dcorr}
\myarrow{65pt}{-46pt}{\siteangle}{-\dcorr}
\myarrow{24.8pt}{-46pt}{\siteangle}{-\dcorr}
}

%% file: editions/pvertex_ten_znzn.tex
\ensuremath{
\begin{matrix}
    \texttransparent{1}{
    \includegraphics[scale=0.108]{images/p4leg_print_crop.pdf}
    }
\end{matrix}
\def \myx {25pt}
\def \myy {25pt}
\def \myxcorr {-0pt} 
\hspace{-\myx}
\raisebox{\myy}{
$\ket{a0{;}\alpha}$
}
\hspace{\myx}
\hspace{\myxcorr}
\def \myx {100pt}
\def \myy {-22pt}
\hspace{-\myx}
\raisebox{\myy}{
$\ket{c0{;}\gamma}$
}
\hspace{\myx}
\hspace{\myxcorr}
\def \myx {90pt}
\def \myy {-9pt}
\hspace{-\myx}
\raisebox{\myy}{
$\ket{b0{;}\beta}$
}
\hspace{\myx}
\hspace{\myxcorr}
\def \myx {165pt}
\def \myy {13pt}
\hspace{-\myx}
\raisebox{\myy}{
$\ket{d0{;}\delta}$
}
\hspace{\myx}
\hspace{\myxcorr}
\def \myx {165pt}
\def \myy {15pt}
\hspace{-\myx}
\raisebox{\myy}{
$V_i$
}
\hspace{\myx}
\def \myx {180pt}
\def \myy {-10pt}
\hspace{-\myx}
\raisebox{\myy}{
$g_i$
}
\hspace{\myx}
\def \myx {169pt}
\def \myy {1.0pt}
\hspace{-\myx}
\raisebox{\myy}{$
\begin{tikzpicture}[line width=1.5pt, color=gray!165]
    \draw [-{Classical TikZ Rightarrow[length=3.5pt]}](0,0) -- (0.05,0);
\end{tikzpicture}
$}
\hspace{\myx}
\def \myx {214pt}
\def \myy {1.0pt}
\hspace{-\myx}
\raisebox{\myy}{$
\begin{tikzpicture}[line width=1.5pt, color=gray!165]
    \draw [-{Classical TikZ Rightarrow[length=3.5pt]}](0,0) -- (0.05,0);
\end{tikzpicture}
$}
\hspace{\myx}
\def \myx {189.5pt}
\def \myy {11.5pt}
\hspace{-\myx}
\raisebox{\myy}{
\rotatebox[origin=c]{218}{$
\begin{tikzpicture}[line width=1.5pt, color=gray!165]
    \draw [-{Classical TikZ Rightarrow[length=3.5pt]}](0,0) -- (0.05,0);
\end{tikzpicture}
$}}
\hspace{\myx}
\def \myx {228.5pt}
\def \myy {-9.5pt}
\hspace{-\myx}
\raisebox{\myy}{
\rotatebox[origin=c]{218}{$
\begin{tikzpicture}[line width=1.5pt, color=gray!165]
    \draw [-{Classical TikZ Rightarrow[length=3.5pt]}](0,0) -- (0.05,0);
\end{tikzpicture}
$}}
\hspace{\myx}
\hspace{-185pt}
}

%% file: editions/qvertex_ten_znzn.tex
\ensuremath{
\begin{matrix}
    \includegraphics[scale=1.1]{images/4leg.pdf}
\end{matrix}
\texttransparent{1}{\llap{$
\begin{matrix}
    \includegraphics[trim= 37.8 0 0 0, scale=1.11]{images/bluesquare.pdf}
\end{matrix}
$}}
\def \myx {25pt}
\def \myy {25pt}
\def \myxcorr {-0pt} 
\hspace{-\myx}
\raisebox{\myy}{
$\ket{0a{;}\alpha}$
}
\hspace{\myx}
\hspace{\myxcorr}
\def \myx {100pt}
\def \myy {-22pt}
\hspace{-\myx}
\raisebox{\myy}{
$\ket{0c{;}\gamma}$
}
\hspace{\myx}
\hspace{\myxcorr}
\def \myx {90pt}
\def \myy {-7pt}
\hspace{-\myx}
\raisebox{\myy}{
$\ket{0b{;}\beta}$
}
\hspace{\myx}
\hspace{\myxcorr}
\def \myx {165pt}
\def \myy {10pt}
\hspace{-\myx}
\raisebox{\myy}{
$\ket{0d{;}\delta}$
}
\hspace{\myx}
\hspace{\myxcorr}
\def \myx {165pt}
\def \myy {15pt}
\hspace{-\myx}
\raisebox{\myy}{
$V_i$
}
\hspace{\myx}
\def \myx {180pt}
\def \myy {-10pt}
\hspace{-\myx}
\raisebox{\myy}{
$g_i$
}
\hspace{\myx}
\def \myx {169pt}
\def \myy {1.0pt}
\hspace{-\myx}
\raisebox{\myy}{$
\begin{tikzpicture}[line width=1.5pt, color=gray!165]
    \draw [-{Classical TikZ Rightarrow[length=3.5pt]}](0,0) -- (0.05,0);
\end{tikzpicture}
$}
\hspace{\myx}
\def \myx {214pt}
\def \myy {1.0pt}
\hspace{-\myx}
\raisebox{\myy}{$
\begin{tikzpicture}[line width=1.5pt, color=gray!165]
    \draw [-{Classical TikZ Rightarrow[length=3.5pt]}](0,0) -- (0.05,0);
\end{tikzpicture}
$}
\hspace{\myx}
\def \myx {189.5pt}
\def \myy {11.5pt}
\hspace{-\myx}
\raisebox{\myy}{
\rotatebox[origin=c]{218}{$
\begin{tikzpicture}[line width=1.5pt, color=gray!165]
    \draw [-{Classical TikZ Rightarrow[length=3.5pt]}](0,0) -- (0.05,0);
\end{tikzpicture}
$}}
\hspace{\myx}
\def \myx {228.5pt}
\def \myy {-9.5pt}
\hspace{-\myx}
\raisebox{\myy}{
\rotatebox[origin=c]{218}{$
\begin{tikzpicture}[line width=1.5pt, color=gray!165]
    \draw [-{Classical TikZ Rightarrow[length=3.5pt]}](0,0) -- (0.05,0);
\end{tikzpicture}
$}}
\hspace{\myx}
\hspace{-185pt}
}

%% file: editions/pqcopy_ten_znzn.tex
\ensuremath{
\begin{matrix}
    \texttransparent{0.93}{
    \includegraphics[scale=0.0175]{images/pq3leg.png}
    }
\end{matrix}
\def \myx {68pt}
\def \myy {-2pt}
\def \myxcorr {-0pt} 
\hspace{-\myx}
\raisebox{\myy}{
$\ket{a0{;}\alpha}$
}
\hspace{\myx}
\hspace{\myxcorr}
\def \myx {49pt}
\def \myy {-2pt}
\hspace{-\myx}
\raisebox{\myy}{
$\ket{0c{;}\gamma}$
}
\hspace{\myx}
\hspace{\myxcorr}
\def \myx {114pt}
\def \myy {-16pt}
\def \myxcorr {-15pt} 
\hspace{-\myx}
\raisebox{\myy}{
$\ket{(-b)b{;}1}$
}
\hspace{\myx}
\hspace{\myxcorr}
\def \myx {127pt}
\def \myy {18pt}
\hspace{-\myx}
\raisebox{\myy}{
$C_l$
}
\hspace{\myx}
\def \myx {125pt}
\def \myy {7.5pt}
\hspace{-\myx}
\raisebox{\myy}{$
\begin{tikzpicture}[line width=1.5pt, color=gray!165]
    \draw [-{Classical TikZ Rightarrow[length=3.5pt]}](0,0) -- (0.05,0);
\end{tikzpicture}
$}
\hspace{\myx}
\def \myx {156pt}
\def \myy {7.5pt}
\hspace{-\myx}
\raisebox{\myy}{$
\begin{tikzpicture}[line width=1.5pt, color=gray!165]
    \draw [-{Classical TikZ Rightarrow[length=3.5pt]}](0,0) -- (0.05,0);
\end{tikzpicture}
$}
\hspace{\myx}
\def \myx {152.1pt}
\def \myy {-2pt}
\hspace{-\myx}
\raisebox{\myy}{
\rotatebox[origin=c]{90}{$
\begin{tikzpicture}[line width=1.5pt, color=gray!165]
    \draw [-{Classical TikZ Rightarrow[length=3.5pt]}](0,0) -- (0.05,0);
\end{tikzpicture}
$}}
\hspace{\myx}
\hspace{-115pt}
}

%% file: editions/qpcopy_ten_znzn.tex
\ensuremath{
\begin{matrix}
    \texttransparent{0.93}{
    \includegraphics[scale=0.0175]{images/qp3leg.png}
    }
\end{matrix}
\def \myx {68pt}
\def \myy {-2pt}
\def \myxcorr {-0pt} 
\hspace{-\myx}
\raisebox{\myy}{
$\ket{0a{;}\alpha}$
}
\hspace{\myx}
\hspace{\myxcorr}
\def \myx {49pt}
\def \myy {-2pt}
\hspace{-\myx}
\raisebox{\myy}{
$\ket{c0{;}\gamma}$
}
\hspace{\myx}
\hspace{\myxcorr}
\def \myx {114pt}
\def \myy {-16pt}
\def \myxcorr {-15pt} 
\hspace{-\myx}
\raisebox{\myy}{
$\ket{b(-b){;}1}$
}
\hspace{\myx}
\hspace{\myxcorr}
\def \myx {127pt}
\def \myy {18pt}
\hspace{-\myx}
\raisebox{\myy}{
$C_l$
}
\hspace{\myx}
\def \myx {124.5pt}
\def \myy {7.5pt}
\hspace{-\myx}
\raisebox{\myy}{$
\begin{tikzpicture}[line width=1.5pt, color=gray!165]
    \draw [-{Classical TikZ Rightarrow[length=3.5pt]}](0,0) -- (0.05,0);
\end{tikzpicture}
$}
\hspace{\myx}
\def \myx {156pt}
\def \myy {7.5pt}
\hspace{-\myx}
\raisebox{\myy}{$
\begin{tikzpicture}[line width=1.5pt, color=gray!165]
    \draw [-{Classical TikZ Rightarrow[length=3.5pt]}](0,0) -- (0.05,0);
\end{tikzpicture}
$}
\hspace{\myx}
\def \myx {152.1pt}
\def \myy {-2pt}
\hspace{-\myx}
\raisebox{\myy}{
\rotatebox[origin=c]{90}{$
\begin{tikzpicture}[line width=1.5pt, color=gray!165]
    \draw [-{Classical TikZ Rightarrow[length=3.5pt]}](0,0) -- (0.05,0);
\end{tikzpicture}
$}}
\hspace{\myx}
\hspace{-115pt}
}

%% file: sections/H_conclusion_luca.tex
\section{Conclusion}
\label{sec:conclusion}

In this work, we have presented a new tensor network ansatz designed to describe states in specific sectors of gauge-invariant theories by explicitly utilizing symmetric elementary tensors. Building on the approach of \cite{tagliacozzo2011}, which identifies symmetric tensor networks with specific instances of duality transformation---a perspective recently extended to the generic scenario in \cite{lootens2023}---our results further suggest the existence of a duality transformation between gauge-invariant systems and globally invariant systems. It will be very important to understand the role of the global symmetry in this new formulation. Indeed, the \zn LGT support  topological phases which get embedded within the larger Hilbert space of  globally symmetric \znzn states, and thus it is important to understand if any feature is left of topological order in these globally symmetric models.

We reviewed how the standard gauge invariant TN can be encoded a finite depth unitary circuit acting on the physical degrees of freedom and additional ancilla degrees of freedom, which must be projected onto the appropriate state at the end of the circuit, in oder to select the desired sector of the gauge-invariant Hilbert space. It would be interesting to see if the present results can also  thus be rephrased in the modern framework of finite-depth circuits combined with measurements, enabling access to \zn topological phases starting from product states.

Also, at this stage, it seems to us that the present formulation goes beyond the use of purely Clifford circuits to construct the \ztwo symmetric ansatz, thus going beyond the new paradigm of Clifford enhanced TN of which the previous ansatz were exemplary issues.

Practically, the new ansatz we present here paves the way to explicitly address specific gauge-invariant sectors of the Hilbert space using globally symmetric tensor networks.
We therefore anticipate that it  will play a crucial role in variational studies of lattice gauge theories on the lattice.

%% file: sections/I_aknowledgments_luca.tex
\begin{acknowledgments}

This work summarizes some of the results from the master’s thesis of MC. The thesis originated with the aim of unifying the approach used in constrained DMRG \cite{chepiga2019} with the methodology  employed in gauge-invariant tensor networks \cite{tagliacozzo2014}. 
We originally realized that the the natural connection is contained in the approach described in \cite{silvi2014}, where the Hilbert space on each link is doubled. Each copy of the link Hilbert space is absorbed into the nearest vertex, and an independent 
$U(1)$ symmetry constraint is imposed on every link of the lattice.

The one-dimensional formulation of this approach is particularly straightforward and facilitated several unpublished numerical simulations, which were carried out in collaboration with Umberto Borla using TenPy. We are grateful to  Umberto for the joint discussions and contributions to these related formulations, and our joint results will appear elsewhere. 

The focus of MC’s thesis was to generalize this approach to two dimensions. However, we quickly realized that in 2D, such an approach would require handling prohibitively large PEPS tensors. Specifically, on a square lattice, each vertex would need to absorb four copies of the link Hilbert space, making the method computationally infeasible in its current form.

While such an approach has been successfully pursued by Montangero's group using tree tensor networks, we realized during the initial stages of MC’s thesis that the PEPS ansatz could remain as simple as the original gauge-invariant ansatz in \cite{tagliacozzo2014} by doubling the symmetries on the links rather than doubling the number of constituents on them. Given the theoretical significance of this observation, we decided to publish these results separately.

 MC was hosted by Jan Von Delft during his thesis and we would like to thank him and his group for the support and the discussions on this topic we had during the thesis.

\begin{sloppypar}
LT acknowledges support from the Proyecto Sinérgico CAM Y2020/TCS-6545 NanoQuCo-CM,
the CSIC Research Platform on Quantum Technologies PTI-001, and from the Grant TED2021-130552B-C22 funded by MCIN/AEI/10.13039/501100011033 and by the ``European Union NextGenerationEU/PRTR'',
and Grant PID2021-127968NB-I00 funded by MCIN/AEI/10.13039/501100011033.
\end{sloppypar}

\end{acknowledgments}